\documentclass[twocolumn]{aastex63}
\usepackage{amsmath}
\usepackage{graphicx}
\usepackage{natbib}
\usepackage[scaled]{helvet}
\usepackage{epsfig}
\usepackage{url}
\bibpunct{(}{)}{;}{a}{}{,}
\usepackage{textcomp}
\usepackage{mathpazo}
\usepackage{xspace}
\usepackage{hyperref}
\usepackage{apjfonts}
\usepackage{mathrsfs}
\usepackage{verbatim}

\usepackage{color}
\usepackage[normalem]{ulem}
\usepackage{multirow}

\newcommand{\bjdtdb}{\ensuremath{\rm {BJD_{TDB}}}}
\newcommand{\feh}{\ensuremath{\left[{\rm Fe}/{\rm H}\right]}}

\newcommand{\teff}{\ensuremath{T_{\rm eff}}\xspace}
\newcommand{\logg}{\ensuremath{\log g}}

\newcommand{\msun}{\ensuremath{\,M_\Sun}}
\newcommand{\rsun}{\ensuremath{\,R_\Sun}}
\newcommand{\lsun}{\ensuremath{\,L_\Sun}}
\newcommand{\mj}{\ensuremath{\,M_{\rm J}}}
\newcommand{\rj}{\ensuremath{\,R_{\rm J}}}

\newcommand{\re}{\ensuremath{\,R_{\rm \Earth}}\xspace}

\newcommand{\fave}{\langle F \rangle}
\newcommand{\fluxcgs}{10$^9$ erg s$^{-1}$ cm$^{-2}$}

\newcommand{\Kepler}{{\it Kepler}}
\newcommand{\Ktwo}{{\it K2}}
\newcommand{\ktwo}{{\it K2}}
\newcommand{\tess}{{\it TESS}}

\newcommand{\rstar}{\ensuremath{R_{*}}}

\newcommand{\be}{\begin{equation}}
\newcommand{\ee}{\end{equation}}

\newcommand{\TESS}{{\it TESS}}

\begin{document}

\title{The \textit{K2} \& \textit{TESS} Synergy I: Updated Ephemerides and Parameters for K2-114, K2-167, K2-237, \& K2-261}

\newcommand{\cfa}{Center for Astrophysics \textbar \ Harvard \& Smithsonian, 60 Garden St, Cambridge, MA 02138, USA}
\newcommand{\umich}{Astronomy Department, University of Michigan, 1085 S University Avenue, Ann Arbor, MI 48109, USA}
\newcommand{\utaustin}{Department of Astronomy, The University of Texas at Austin, Austin, TX 78712, USA}
\newcommand{\MIT}{Department of Physics and Kavli Institute for Astrophysics and Space Research, Massachusetts Institute of Technology, Cambridge, MA 02139, USA}
\newcommand{\MITEPS}{Department of Earth, Atmospheric and Planetary Sciences, Massachusetts Institute of Technology,  Cambridge,  MA 02139, USA}
\newcommand{\uflorida}{Department of Astronomy, University of Florida, 211 Bryant Space Science Center, Gainesville, FL, 32611, USA}
\newcommand{\riverside}{Department of Earth and Planetary Sciences, University of California, Riverside, CA 92521, USA}
\newcommand{\usq}{University of Southern Queensland, West St, Darling Heights QLD 4350, Australia}
\newcommand{\ames}{NASA Ames Research Center, Moffett Field, CA, 94035, USA}
\newcommand{\geneva}{Observatoire de l’Universit\'e de Gen\`eve, 51 chemin des Maillettes, 1290 Versoix, Switzerland}
\newcommand{\uw}{Astronomy Department, University of Washington, Seattle, WA 98195 USA}
\newcommand{\warwick}{Department of Physics, University of Warwick, Gibbet Hill Road, Coventry CV4 7AL, UK}
\newcommand{\warwickceh}{Centre for Exoplanets and Habitability, University of Warwick, Gibbet Hill Road, Coventry CV4 7AL, UK}
\newcommand{\princeton}{Department of Astrophysical Sciences, Princeton University, 4 Ivy Lane, Princeton, NJ, 08544, USA}
\newcommand{\liege}{Space Sciences, Technologies and Astrophysics Research (STAR) Institute, Universit\'e de Li\`ege, 19C All\'ee du 6 Ao\^ut, 4000 Li\`ege, Belgium}
\newcommand{\vanderbilt}{Department of Physics and Astronomy, Vanderbilt University, Nashville, TN 37235, USA}
\newcommand{\fisk}{Department of Physics, Fisk University, 1000 17th Avenue North, Nashville, TN 37208, USA}
\newcommand{\columbia}{Department of Astronomy, Columbia University, 550 West 120th Street, New York, NY 10027, USA}
\newcommand{\toronto}{Dunlap Institute for Astronomy and Astrophysics, University of Toronto, Ontario M5S 3H4, Canada}
\newcommand{\unc}{Department of Physics and Astronomy, University of North Carolina at Chapel Hill, Chapel Hill, NC 27599, USA}
\newcommand{\iac}{Instituto de Astrof\'isica de Canarias (IAC), E-38205 La Laguna, Tenerife, Spain}
\newcommand{\lalaguna}{Departamento de Astrof\'isica, Universidad de La Laguna (ULL), E-38206 La Laguna, Tenerife, Spain}
\newcommand{\louisville}{Department of Physics and Astronomy, University of Louisville, Louisville, KY 40292, USA}
\newcommand{\aavso}{American Association of Variable Star Observers, 49 Bay State Road, Cambridge, MA 02138, USA}
\newcommand{\utokyo}{The University of Tokyo, 7-3-1 Hongo, Bunky\={o}, Tokyo 113-8654, Japan}
\newcommand{\naoj}{National Astronomical Observatory of Japan, 2-21-1 Osawa, Mitaka, Tokyo 181-8588, Japan}
\newcommand{\jstpresto}{JST, PRESTO, 7-3-1 Hongo, Bunkyo-ku, Tokyo 113-0033, Japan}
\newcommand{\astrobiojapan}{Astrobiology Center, 2-21-1 Osawa, Mitaka, Tokyo 181-8588, Japan}
\newcommand{\ctio}{Cerro Tololo Inter-American Observatory, Casilla 603, La Serena, Chile}
\newcommand{\nexsci}{Caltech IPAC -- NASA Exoplanet Science Institute 1200 E. California Ave, Pasadena, CA 91125, USA}
\newcommand{\ucsc}{Department of Astronomy and Astrophysics, University of
California, Santa Cruz, CA 95064, USA}
\newcommand{\gsfc}{Exoplanets and Stellar Astrophysics Laboratory, Code 667, NASA Goddard Space Flight Center, Greenbelt, MD 20771, USA}
\newcommand{\sgtinc}{SGT, Inc./NASA AMES Research Center, Mailstop 269-3, Bldg T35C, P.O. Box 1, Moffett Field, CA 94035, USA}
\newcommand{\chile}{Center of Astro-Engineering UC, Pontificia Universidad Cat\'olica de Chile, Av. Vicu\~{n}a Mackenna 4860, 7820436 Macul, Santiago, Chile}
\newcommand{\Pontificia}{Instituto de Astrof\'isica, Pontificia Universidad Cat\'olica de Chile, Av.\ Vicu\~na Mackenna 4860, Macul, Santiago, Chile}
\newcommand{\Millennium}{Millennium Institute for Astrophysics, Chile}
\newcommand{\maxplank}{Max-Planck-Institut f\"ur Astronomie, K\"onigstuhl 17, Heidelberg 69117, Germany}
\newcommand{\utdallas}{Department of Physics, The University of Texas at Dallas, 800 West
Campbell Road, Richardson, TX 75080-3021 USA}
\newcommand{\MauryLewin}{Maury Lewin Astronomical Observatory, Glendora, CA 91741, USA}
\newcommand{\umbc}{University of Maryland, Baltimore County, 1000 Hilltop Circle, Baltimore, MD 21250, USA}
\newcommand{\osu}{Department of Astronomy, The Ohio State University, 140 West 18th Avenue, Columbus, OH 43210, USA}
\newcommand{\MITAA}{Department of Aeronautics and Astronautics, MIT, 77 Massachusetts Avenue, Cambridge, MA 02139, USA}
\newcommand{\openu}{School of Physical Sciences, The Open University, Milton Keynes MK7 6AA, UK}
\newcommand{\swarthmore}{Department of Physics and Astronomy, Swarthmore College, Swarthmore, PA 19081, USA}
\newcommand{\seti}{SETI Institute, Mountain View, CA 94043, USA}
\newcommand{\lehigh}{Department of Physics, Lehigh University, 16 Memorial Drive East, Bethlehem, PA 18015, USA}
\newcommand{\utah}{Department of Physics and Astronomy, University of Utah, 115 South 1400 East, Salt Lake City, UT 84112, USA}
\newcommand{\USNA}{Department of Physics, United States Naval Academy, 572C Holloway Rd., Annapolis, MD 21402, USA}
\newcommand{\DTM}{Department of Terrestrial Magnetism, Carnegie Institution for Science, 5241 Broad Branch Road, NW, Washington, DC 20015, USA}
\newcommand{\UPenn}{The University of Pennsylvania, Department of Physics and Astronomy, Philadelphia, PA, 19104, USA}
\newcommand{\montana}{Department of Physics and Astronomy, University of Montana, 32 Campus Drive, No. 1080, Missoula, MT 59812 USA}
\newcommand{\psu}{Department of Astronomy \& Astrophysics, The Pennsylvania State University, 525 Davey Lab, University Park, PA 16802, USA}
\newcommand{\psust}{Center for Exoplanets and Habitable Worlds, The Pennsylvania State University, 525 Davey Lab, University Park, PA 16802, USA}
\newcommand{\Kutztown}{Department of Physical Sciences, Kutztown University, Kutztown, PA 19530, USA}
\newcommand{\udel}{Department of Physics \& Astronomy, University of Delaware, Newark, DE 19716, USA}
\newcommand{\Westminster}{Department of Physics, Westminster College, New Wilmington, PA 16172}
\newcommand{\steward}{Department of Astronomy and Steward Observatory, University of Arizona, Tucson, AZ 85721, USA}
\newcommand{\saao}{South African Astronomical Observatory, PO Box 9, Observatory, 7935, Cape Town, South Africa}
\newcommand{\salt}{Southern African Large Telescope, PO Box 9, Observatory, 7935, Cape Town, South Africa}
\newcommand{\ssl}{Societ\`{a} Astronomica Lunae, Italy}
\newcommand{\spot}{Spot Observatory, Nashville, TN 37206, USA}
\newcommand{\txamGP}{George P.\ and Cynthia Woods Mitchell Institute for Fundamental Physics and Astronomy, Texas A\&M University, College Station, TX77843 USA}
\newcommand{\txam}{Department of Physics and Astronomy, Texas A\&M university, College Station, TX 77843 USA}
\newcommand{\wellesley}{Department of Astronomy, Wellesley College, Wellesley, MA 02481, USA}
\newcommand{\byu}{Department of Physics and Astronomy, Brigham Young University, Provo, UT 84602, USA}
\newcommand{\Hazelwood}{Hazelwood Observatory, Churchill, Victoria, Australia}
\newcommand{\pest}{Perth Exoplanet Survey Telescope}
\newcommand{\Winer}{Winer Observatory, PO Box 797, Sonoita, AZ 85637, USA}
\newcommand{\icpo}{Ivan Curtis Private Observatory}
\newcommand{\elsauce}{El Sauce Observatory, Chile}
\newcommand{\crow}{Atalaia Group \& CROW Observatory, Portalegre, Portugal}
\newcommand{\dfus}{Dipartimento di Fisica ``E.R.Caianiello'', Universit\`a di Salerno, Via Giovanni Paolo II 132, Fisciano 84084, Italy}
\newcommand{\indfn}{Istituto Nazionale di Fisica Nucleare, Napoli, Italy}
\newcommand{\sotes}{Gabriel Murawski Private Observatory (SOTES)}
\newcommand{\unmex}{Department of Physics \& Astronomy, University of New Mexico, 1919 Lomas Blvd NE, Albuquerque, NM 87131, USA}
\newcommand{\protologic}{Proto-Logic Consulting LLC, Washington, DC 20009, USA}
\newcommand{\gsfcsellers}{GSFC Sellers Exoplanet Environments Collaboration, NASA Goddard Space Flight Center, Greenbelt, MD 20771, USA }
\newcommand{\colorado}{Department of Astrophysical and Planetary Sciences, University of Colorado, Boulder, CO 80309, USA}
\newcommand{\torres}{\altaffiliation{Juan Carlos Torres Fellow}}
\newcommand{\sagan}{\altaffiliation{NASA Sagan Fellow}}
\newcommand{\bernoulli}{\altaffiliation{Bernoulli fellow}}
\newcommand{\gruber}{\altaffiliation{Gruber fellow}}
\newcommand{\kavli}{\altaffiliation{Kavli Fellow}}
\newcommand{\peg}{\altaffiliation{51 Pegasi b Fellow}}
\newcommand{\pappalardo}{\altaffiliation{Pappalardo Fellow}}
\newcommand{\hubble}{\altaffiliation{NASA Hubble Fellow}}

\correspondingauthor{Mma Ikwut-Ukwa} 
\email{mma.ikwut-ukwa@cfa.harvard.edu}

\author[0000-0002-4404-5505]{Mma Ikwut-Ukwa} 
\affiliation{\cfa}

\author[0000-0001-8812-0565]{Joseph E. Rodriguez} 
\affiliation{\cfa}

\author[0000-0001-6637-5401]{Allyson Bieryla} 
\affiliation{\cfa}

\author[0000-0001-7246-5438]{Andrew Vanderburg}
\sagan
\affiliation{\utaustin}

\author[0000-0003-4603-556X]{Teo Mocnik}
\affiliation{\riverside}

\author[0000-0002-7084-0529]{Stephen R.\ Kane}
\affiliation{\riverside}

\author[0000-0002-8964-8377]{Samuel N. Quinn} 
\affiliation{\cfa}

\author[0000-0001-8020-7121]{Knicole D.\ Col\'on} 
\affiliation{\gsfc}
\affiliation{\gsfcsellers}

\author[0000-0002-4891-3517]{George Zhou} 
\hubble
\affiliation{\cfa}

\author[0000-0003-3773-5142]{Jason D. Eastman} 
\affiliation{\cfa}

\author[0000-0003-0918-7484]{Chelsea X. Huang} 
\torres
\affiliation{\MIT}

\author[0000-0001-9911-7388]{David W. Latham} 
\affiliation{\cfa}

\author{Jessie Dotson} 
\affiliation{\ames}

\author[0000-0002-4715-9460]{Jon M. Jenkins} 
\affiliation{\ames}

\author{George R. Ricker} 
\affiliation{\MIT}

\author[0000-0002-6892-6948]{Sara Seager} 
\affiliation{\MIT}
\affiliation{\MITEPS}
\affiliation{\MITAA}

\author[0000-0001-6763-6562]{Roland K. Vanderspek} 
\affiliation{\MIT}

\author[0000-0002-4265-047X]{Joshua N. Winn} 
\affiliation{\princeton}


\author[0000-0001-7139-2724]{Thomas Barclay} 
\affiliation{\gsfc}
\affiliation{\umbc}
\affiliation{\gsfcsellers}

\author[0000-0002-3306-3484]{Geert Barentsen}
\affiliation{\ames}

\author[0000-0002-3321-4924]{Zachory Berta-Thompson} 
\affiliation{\colorado}

\author[0000-0002-9003-484X]{David Charbonneau}
\affiliation{\cfa}

\author[0000-0003-2313-467X]{Diana Dragomir} 
\affiliation{\unmex}

\author[0000-0002-6939-9211]{Tansu Daylan} 
\kavli
\affiliation{\MIT}


\author[0000-0002-3164-9086]{Maximilian~N.~G{\"u}nther}  
\torres
\affiliation{\MIT}

\author[0000-0002-3385-8391]{Christina Hedges} 
\affiliation{\ames}

\author{Christopher E. Henze}
\affiliation{\ames}

\author{Scott McDermott} 
\affiliation{\protologic}

\author{Joshua E. Schlieder} 
\affiliation{\gsfc}
\affiliation{\gsfcsellers}

\author[0000-0003-1309-2904]{Elisa V. Quintana} 
\affiliation{\gsfc}
\affiliation{\gsfcsellers}

\author[0000-0002-6148-7903]{Jeffrey C. Smith} 
\affiliation{\seti}
\affiliation{\ames}

\author[0000-0002-6778-7552]{Joseph D. Twicken} 
\affiliation{\seti}
\affiliation{\ames}

\author[0000-0003-4755-584X]{Daniel A. Yahalomi} 
\affiliation{\cfa}

\shorttitle{The \ktwo\ \& \TESS\ Synergy I}
\shortauthors{Ikwut-Ukwa et al.}

\begin{abstract}
{Although the Transiting Exoplanet Survey Satellite ({\it TESS}) primary mission observed the northern and southern ecliptic hemispheres, generally avoiding the ecliptic, and the {\it Kepler} space telescope during the {\it K2} mission could only observe near the ecliptic, many of the {\it K2} fields extend far enough from the ecliptic plane that sections overlap with {\it TESS} fields.} Using photometric observations from both {\it K2} and {\it TESS}, combined with archival spectroscopic observations, we globally modeled four known planetary systems discovered by {\it K2} that were observed in the first year of the primary {\it TESS} mission. Specifically, we provide updated ephemerides and system parameters for K2-114 b, K2-167 b, K2-237 b, and K2-261 b. These were some of the first {\it K2} planets to be observed by {\it TESS} in the first year and include three Jovian sized planets and a sub-Neptune with orbital periods less than 12 days. In each case, the updated ephemeris significantly reduces the uncertainty in prediction of future times of transit, which is valuable for planning observations with the James Webb Space Telescope and other future facilities. The {\it TESS} extended mission is expected to observe about half of the {\it K2} fields, providing the opportunity to perform this type of analysis on a larger number of systems. 
\end{abstract}

\section{Introduction}
\begin{table*}
\scriptsize
\setlength{\tabcolsep}{2pt}
\centering
\caption{Literature Properties for K2-114, K2-167, K2-237, \& K2-261 \label{tab:LitProps}}
\begin{tabular}{llcccccc}
Parameter & Description &K2-114&K2-167&K2-237 & K2-261&Source\\
\hline 
Other identifiers& & 2MASS J08313191+1155202 & HD 212657 & 2MASS J16550453-2842380& TYC 255-257-1\\
& & TOI-514 & TOI-1407 & TOI-1049 & TOI-685 \\
& & TIC 366576758 & TIC 69747919 & TIC 16288184 & TIC 281731203 \\
& & EPIC 211418729 & EPIC 205904628 & EPIC 229426032 & EPIC 201498078\\
& &  &  \\
$\alpha_{J2000}$\dotfill	&Right Ascension (RA)\dotfill & 08:31:31.913 & 22:26:18.190 &16:55:04.534&10:52:07.779&  1	\\
$\delta_{J2000}$\dotfill	&Declination (Dec)\dotfill & +11:55:20.156& -18:00:40.220 &-28:42:38.0150&+00:29:36.086& 1	\\
\hline 
\\
$l$\dotfill     & Galactic Longitude\dotfill & 127.88291401$^\circ$ & 336.576123339$^\circ$ &253.76884794$^\circ$&163.03231188$^\circ$& 1\\
$b$\dotfill     & Galactic Latitude\dotfill & 11.92225517$^\circ$ & -18.01166503$^\circ$ &	-28.71058359$^\circ$&+00.49316696$^\circ$& 1\\
\\
B$_T$\dotfill	&Tycho B$_T$ mag.\dotfill & --- &  8.879 $\pm$ 0.02	&---&11.805$\pm$0.086& 2	\\
V$_T$\dotfill	&Tycho V$_T$ mag.\dotfill & ---	&  8.301 $\pm$ 0.02	&---&10.717$\pm$0.059& 2	\\
${\rm G}$\dotfill   & Gaia $G$ mag.\dotfill &14.2751$\pm$0.02&8.104$\pm$0.02&11.467$\pm$0.02&10.459$\pm$0.02& 1\\
${\rm G Bp}$\dotfill   & Gaia $B_P$ mag.\dotfill &14.806$\pm$0.02&8.401$\pm$0.02&11.776$\pm$0.02&10.872$\pm$0.02& 1\\
${\rm G Rp}$\dotfill   & Gaia $R_P$ mag.\dotfill &13.615$\pm$0.02&7.688$\pm$0.02&11.012$\pm$0.02&9.917$\pm$0.02& 1\\
\\
J\dotfill			& 2MASS J mag.\dotfill & 12.835 $\pm$ 0.02 &7.202$\pm$0.02&10.508$\pm$0.02&9.337$\pm$0.03& 3	\\
H\dotfill			& 2MASS H mag.\dotfill & 12.386 $\pm$ 0.03 &6.974$\pm$0.04&10.268$\pm$0.02&8.920$\pm$0.04& 3	\\
K$_S$\dotfill			& 2MASS ${\rm K_S}$ mag.\dotfill & 12.304 $\pm$ 0.03& 6.887$\pm$0.03&10.217$\pm$&8.890$\pm$0.02&  3	\\
\\
\textit{WISE1}\dotfill		& \textit{WISE1} mag.\dotfill & 9.213 $\pm$ 0.022 & 6.810$\pm$0.055 &10.105$^{0.03}_{0.023}$&8.828$^{0.03}_{0.023}$&  4	\\
\textit{WISE2}\dotfill		& \textit{WISE2} mag.\dotfill & 9.245 $\pm$ 0.02 &6.866$^{0.03}_{0.02}$ &10.129$^{0.03}_{0.02}$&8.897$^{0.03}_{0.02}$& 4 \\
\textit{WISE3}\dotfill		& \textit{WISE3} mag.\dotfill & ---& 6.906$^{0.03}_{0.017}$ &9.972$\pm$0.077&8.819$\pm$0.031&4	\\
\textit{WISE4}\dotfill		& \textit{WISE4} mag.\dotfill & ---& 6.917$\pm$0.1 &---&---& 4	\\
\\
$\mu_{\alpha}$\dotfill		& Gaia proper motion\dotfill & -13.149$\pm$0.061 &73.606$\pm$0.105&-8.568$\pm$0.100&-23.664$\pm$0.075& 1 \\
                    & \hspace{3pt} in RA (mas yr$^{-1}$)	&  \\
$\mu_{\delta}$\dotfill		& Gaia proper motion\dotfill 	&  -2.452$\pm$0.037 &-114.505$\pm$0.093	&-5.562$\pm$0.055&-44.171$\pm$0.068&  1 \\
                    & \hspace{3pt} in DEC (mas yr$^{-1}$) &  \\
$\pi^\dagger$\dotfill & Gaia Parallax (mas) \dotfill & 2.1554$\pm$  0.0485 & 12.4148$\pm$0.0786 &3.23058$\pm$0.0779&4.74218$\pm$0.05398&  1 \\
\hline
\end{tabular}
\begin{flushleft}
 \footnotesize{ \textbf{\textsc{NOTES:}}
 The uncertainties of the photometry have a systematic error floor applied. \\
 $\ddagger$ RA and Dec are in epoch J2000. The coordinates come from Vizier where the Gaia RA and Dec have been precessed to J2000 from epoch J2015.5.\\
 $\dagger$ Parallaxes here are corrected for the 82 $\mu$as offset reported in \citet{Stassun:2018}.\\
 References: $^1$\citet{Gaia:2018}, $^2$\citet{Hog:2000}, $^3$\citet{Cutri:2003}, $^4$\citet{Zacharias:2017}
}
\end{flushleft}
\end{table*}

The upcoming generation of telescopes, including the James Webb Space Telescope ({\it JWST}, \citealp{Gardner:2006}) and the Extremely Large Telescopes (ELTs) with highly sensitive instrumentation \citep[e.g.][]{Szentgyorgyi:2016}, will revolutionize the study of exoplanets. These telescopes will enable high-precision follow-up observations of transiting exoplanets, including atmospheric characterization of small planets (R$_P<4\re$). Additionally, future missions are being planned with the hope of detecting biosignatures in the atmospheres of small planets \citep{Roberge:2018, Gaudi:2018}. The targets for these new telescopes will be planets previously discovered by missions like NASA's \textit{Kepler} \citep{Borucki:2010} and the Transiting Exoplanet Survey Satellite (\TESS, \citealp{Ricker:2015}), as well as ground-based transit surveys like WASP \citep{Butters:2010}, HATNET \citep{Bakos:2010}, KELT \citep{Pepper:2007, Pepper:2012}, MEarth \citep{Irwin:2015, Dittmann:2017a}, and TRAPPIST \citep{Gillon:2011}, and more recently, SPECULOOS \citep{Delrez:2018} and NGTS \citep{Wheatley:2018}. These future facilities will require efficient scheduling, meaning the transit times predicted for exoplanet targets will need to be both accurate and precise. The high cost of operations for {\it JWST} particularly will necessitate precise ephemerides in order to use resources efficiently. Currently, the predicted transit times of many previously discovered planets (when projected through the {\it JWST} era) have ephemerides too imprecise to meet these demands.

\begin{figure*}[ht!]
	\centering\vspace{.0in}
	\includegraphics[width=0.8\linewidth, clip]{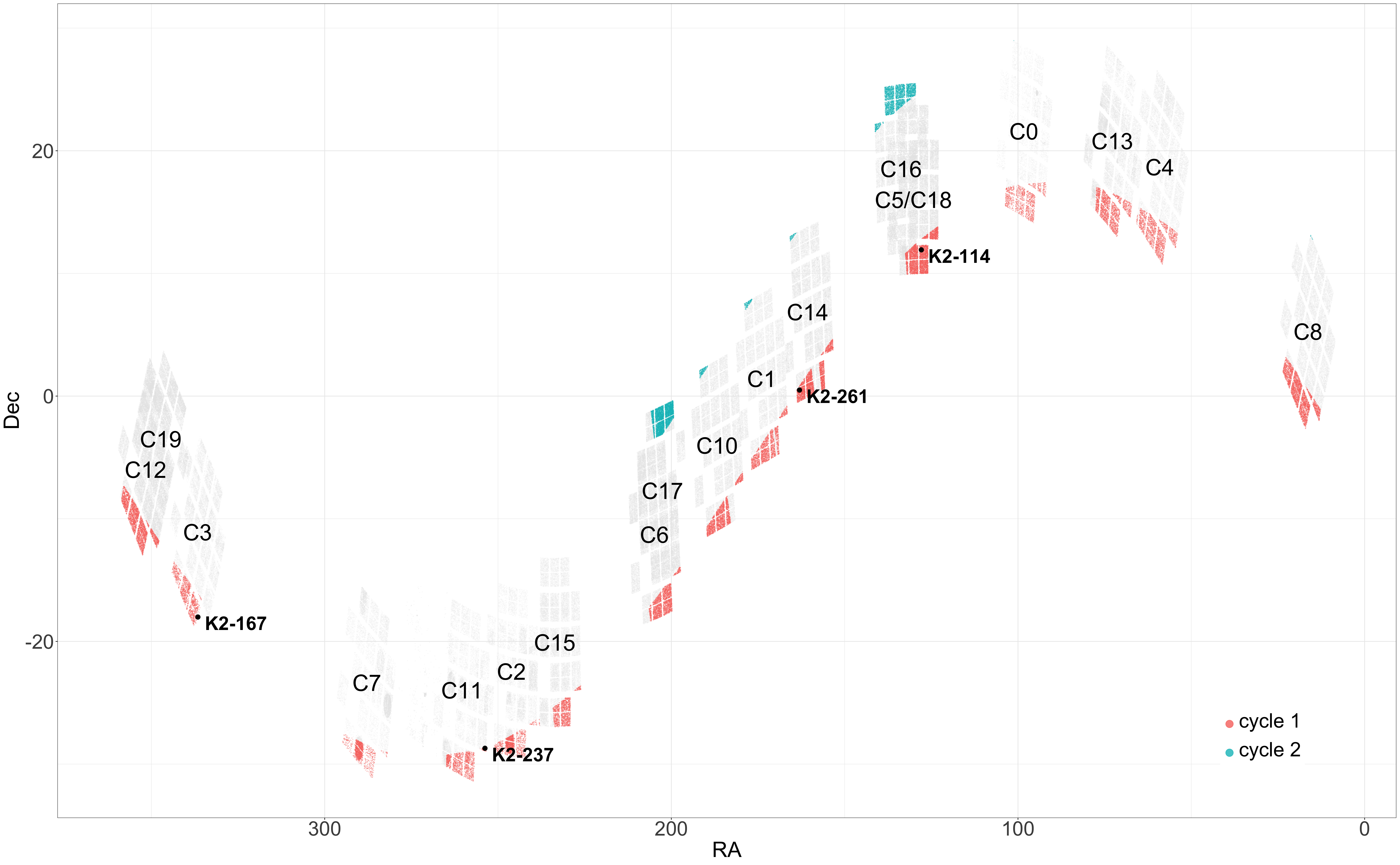}\\		\caption{Map of \ktwo\ targets that were observed in  \TESS\ years 1 (red) and 2 (blue).}
	\label{fig:Map} 
\end{figure*}

\par The \textit{Kepler} space telescope, NASA's first mission aimed at discovering transiting exoplanets, led to the discovery of over {2300 planets} and the identification of thousands more candidates \citep{Thompson:2018}.\footnote{\label{archive}\url{exoplanetarchive.ipac.caltech.edu}} However, by the end of the primary mission in early 2013, a second reaction wheel on the spacecraft failed, compromising the spacecraft's ability to point. The \ktwo\ mission \citep{Howell:2014} solved the spacecraft's pointing ability by balancing {solar radiation pressure} to stabilize the \textit{Kepler} spacecraft. This led to a survey of the ecliptic plane, providing another opportunity to discover planets around bright, nearby stars. Before being retired in 2018, \ktwo\ completed 18 full observing campaigns of approximately 80 days, discovering over 400 additional planets.\textsuperscript{\ref{archive}} Many of the planets discovered by \ktwo\ now have stale ephemerides, since some were observed as early as 2014 and the first was announced in December of 2014 \citep{Vanderburg:2015}. This hinders our ability to precisely predict upcoming times of transit. {Previous efforts addressing this issue have used follow up transit observations, such as from NASA's Spitzer Space Telescope, to refine K2 ephemerides \citep[e.g.][]{Benneke:2017,Livingston:2019}.}

\par The Transiting Exoplanet Survey Satellite (\TESS) provides an opportunity to update the ephemerides for many more \ktwo\ planets and also to improve the stellar and planetary parameters. \TESS\ launched in April 2018 with the goal of discovering thousands of new planets around nearby, bright stars \citep{Ricker:2015}. Now in the second year of its primary mission, \TESS\ has so far discovered {51 planets appearing in the refereed literature}\textsuperscript{\ref{archive}} \footnote{as of June 4, 2020} \citep[e.g.][]{Huang:2018, Vanderspek:2019, Rodriguez:2019} and over a thousand planet candidates\footnote{\url{exofop.ipac.caltech.edu/tess}} (N. Guerrero et al. \textit{submitted}). Although \TESS's on-sky footprint avoids the ecliptic plane in its primary mission, the \ktwo\ fields extend far enough out of the ecliptic to partially overlap with the \TESS\ fields, and for \ktwo\ campaign 19 it was simultaneous \citep{Barclay:2018b}. {\citet{Dotson:2020} compared the K2 target list with the planned TESS observations using the TESS Visibility Tool\footnote{\url{heasarc.gsfc.nasa.gov/cgi-bin/tess/webtess/wtv.py}} and concluded that during Cycle 1 (the first year) of the primary mission, \TESS\ observed 39,451 \ktwo\ targets. By the end of Cycle 2, \TESS\ will have observed a total of 48,633 \ktwo\ targets (see Figure \ref{fig:Map}), and in the first approved extended mission, it will observe over half of the \ktwo\ footprint \citep{Dotson:2020}.\footnote{\url{heasarc.gsfc.nasa.gov/docs/tess/announcement-of-the-tess-extended-mission.html}}} Additionally, the \tess\ ephemerides, specifically systems with only a $\sim$27 day baseline, will degrade in a similar manner as \ktwo\ systems, and will also require future follow up observations to update the predicted times of transit \citep{Dragomir:2020}.

\par In this paper we present a case study of the potential improvement in precision of future transits for \ktwo\ planets observed by \TESS. Using observations from both missions, we can significantly improve the precision and accuracy of the ephemeris for known planetary systems discovered by \ktwo. Additionally, the combined \ktwo\ and \TESS\ data {provide} the opportunity to potentially discover new planets in these systems and has already aided in the vetting of \TESS\ planet candidates. We jointly fit the \TESS\ and \ktwo\ data, with archival radial velocities, to provide updated ephemerides and system parameters for four planetary systems discovered by \ktwo: K2-114, K2-167, K2-237, and K2-261 {\citep{Shporer:2017,Johnson:2018,Livingston:2019,Mayo:2018,Soto:2018,Smith:2019}}. These four systems were some of the first K2 targets to be observed by \TESS\, and were chosen because they were clearly detected in \TESS. Using our analysis, we significantly improve the precision of predicted transit times as projected through the {\it JWST} era, in some cases by an order of magnitude.

\section{Observations and Archival Data}
\label{Obs}
In this section, we discuss the observations used in our analysis to refine and improve the ephemerides and system parameters for future follow-up efforts. See Table \ref{tab:LitProps} for the literature kinematics and magnitudes for K2-114, K2-167, K2-237, and K2-261.

\begin{table*}
\centering\scriptsize
\caption{The dates of the \TESS\ and \ktwo\ observations.}
\begin{tabular}{cccccccc}
\hline
\hline
Target & K2 Campaign & K2 Dates (UT) & TESS Sector & TESS Dates (UT)\\
K2-114 & 5 & 2015 April 27 to July 10 & 7 & 2019 January 07 to February 02 \\
       & 18 & 2018 May 13 to July 2 & & \\
K2-167 & 3 & 2014 November 17 to 2015 January 23 & 2 & 2018 August 22 to September 20\\
K2-237 & 11 & 2016 September 26 to December 07 & 12 & 2019 May 21 to UT 2019 June 19\\
K2-261 & 14 & 2017 June 02 to August 19 & 9 & 2019 February 28 to March 26\\
\hline\\
\end{tabular}
\label{tab:observations}
 \begin{flushleft} 
  \footnotesize{ }
 \end{flushleft}
\end{table*}

\subsection{{\it K2} Photometry}
\label{sec:k2}
\begin{figure*}[!ht]
	\centering\includegraphics[width=0.45\linewidth]{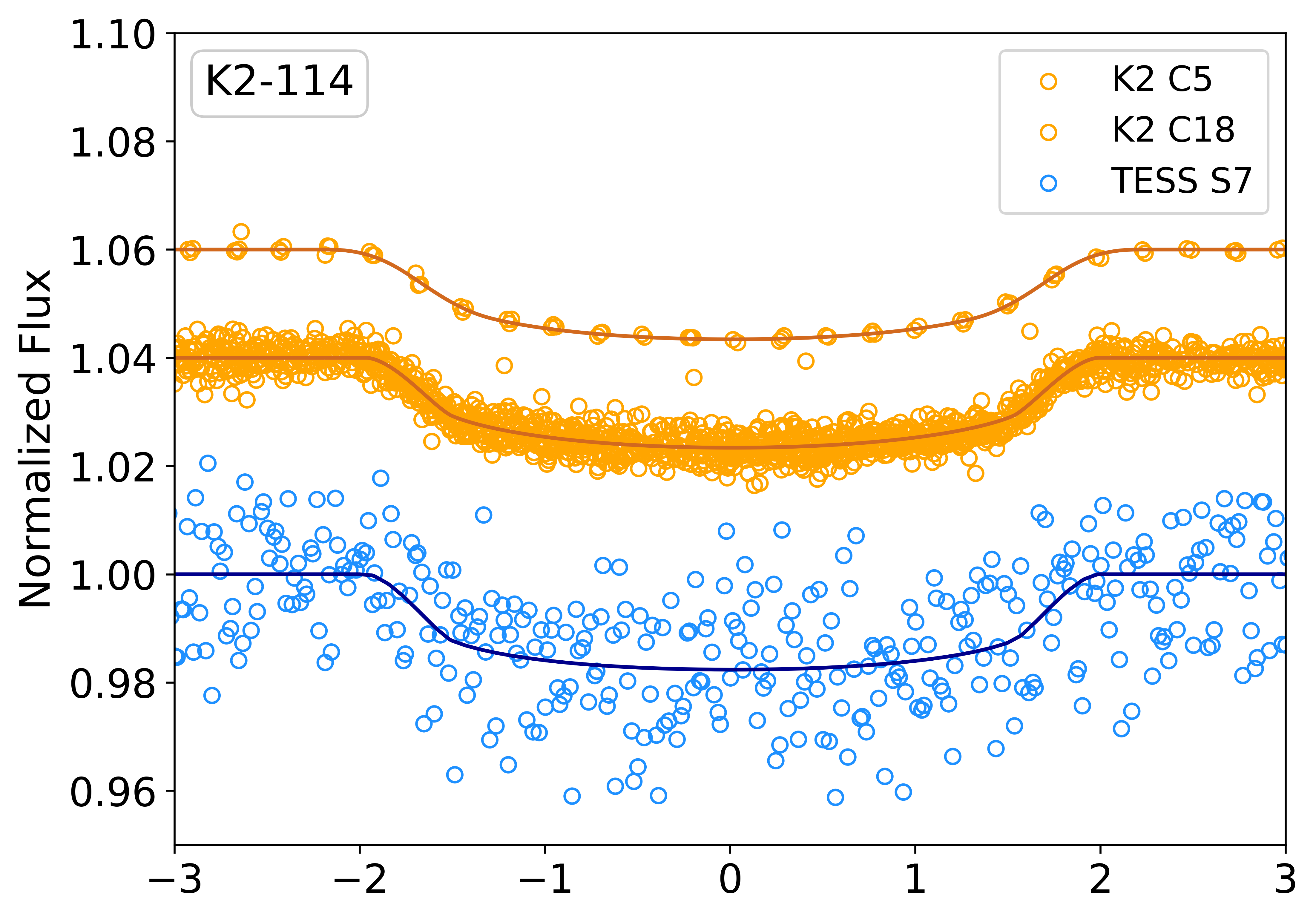}\hspace*{0.5cm}\includegraphics[width=0.45\linewidth]{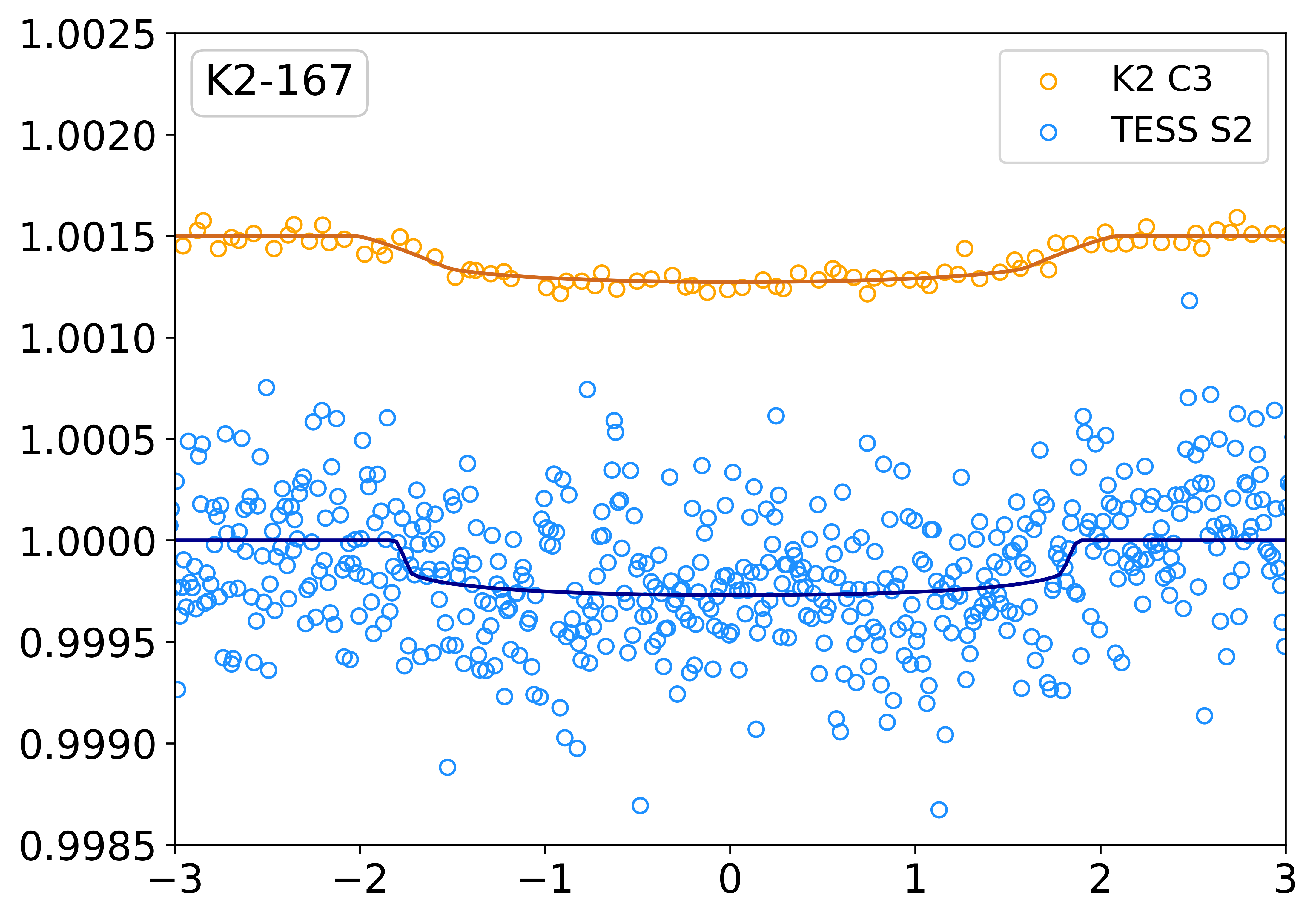}
	\vspace*{0.25cm}
   \centering\includegraphics[width=0.45\linewidth]{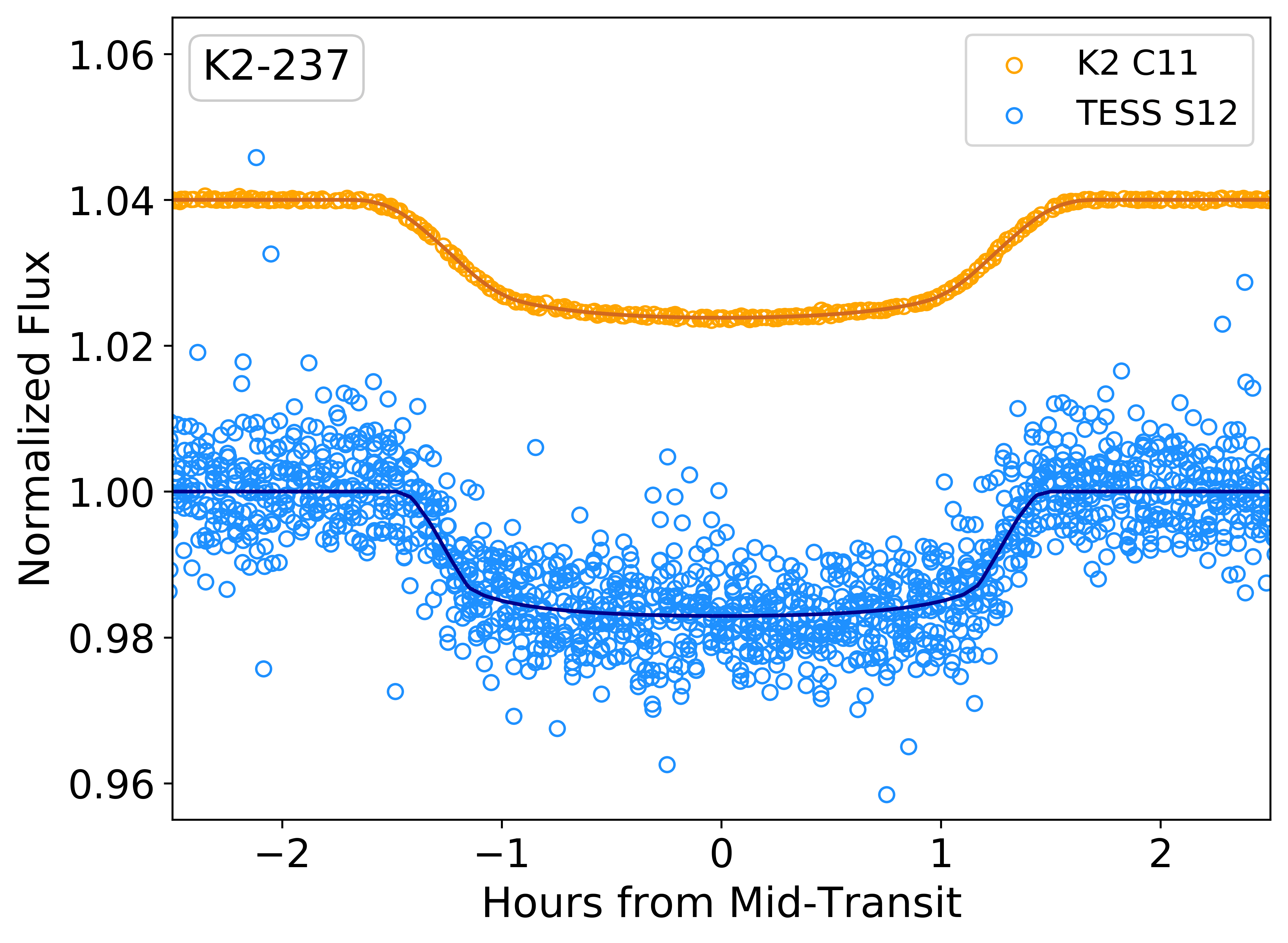}\hspace*{0.5cm}\includegraphics[width=0.45\linewidth]{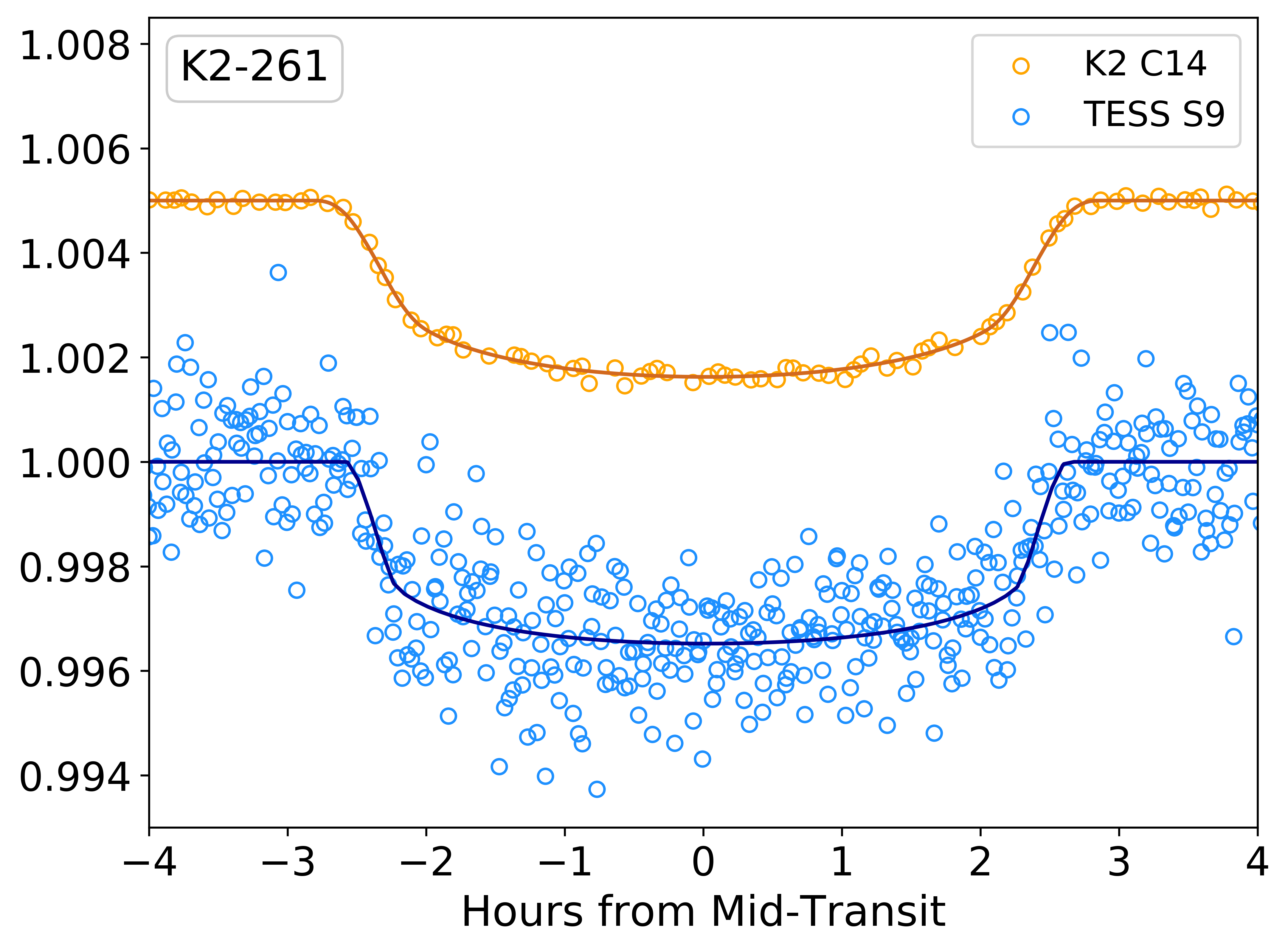}
	\caption{{The phase folded (blue) \tess\ and (gold) \ktwo\ transits} for (top-left) K2-114 b, (top-right) K2-167 b, (bottom-left) K2-237 b, and (bottom-right) K2-261 b. The legend indicates that the \Ktwo\ campaign and \TESS\ sector the target was observed in. The solid color line on each transit represents the best-fit transit model from our EXOFASTv2 global fit (see Section \ref{sec:GlobalModel}). A vertical offset has been applied to the {\it K2} data in each system for visual clarity. K2-114 b was observed at 30-minute cadence in \ktwo\ campaign 5 and then reobserved at 1-minute cadence in campaign 18.}
	\label{fig:transits} 
\end{figure*}

During its lifetime, \Ktwo\ achieved similar precision (after applying corrections) to that of the original four year {\it Kepler} prime mission \citep{Vanderburg:2016b}. For each target, we extracted the light curve from the target pixel files, calibrated by the {\it Kepler} pipeline \citep{Jenkins:2010} and accessed through the {Mikulski Archive for Space Telescopes (MAST).\footnote{\url{mast.stsci.edu/portal/Mashup/Clients/Mast/Portal.html}}} We followed the technique described in \citet{Vanderburg:2014} and \citet{Vanderburg:2016a} to reprocess the light curve, fitting the known planet transit while simultaneously removing known \ktwo\ systematics from spacecraft motion and fitting variability induced by the host star. {We used the default photometric apertures chosen by the pipeline for all planets except for K2-237, where we chose smaller apertures to reduce contamination from nearby stars (see Section \ref{sec:Variability}). We applied a correction to the K2-237 light curve to account for the remaining contaminating light we could not avoid using the measured \Kepler\ pixel response function \citep{Bryson:2010} and the \Kepler\ band magnitudes of nearby stars to calculate the expected flux contamination. The other three stars are sufficiently isolated that the dilution corrections are negligible. } We then flattened the light curve, removing the stellar variability with a spline with break points every 0.75 days. {The K2 phase-folded transit light curves are shown in gold in Figure \ref{fig:transits}.} For the global fitting in Section \ref{sec:GlobalModel}, we used a baseline of one transit duration on either side of the full transit, removing the remaining out-of-transit data. {Each \Ktwo\ target was observed in 30-minute cadence, with the exception of the C18 observations of K2-114, which were taken at 1-minute cadence.} See Table \ref{tab:observations} for the times and campaigns for each target.


\subsection{{\it TESS} Photometry}
\label{sec:TESS}

All four of the \ktwo\ systems (K2-114, K2-167, K2-237, \& K2-261) were pre-selected for two-minute cadence observations by \TESS\ (K2-114 b was a Guest Investigator (GI) target, G011183 PI Kane). Each system was observed by Camera 1 during one of \TESS' $\sim$27-day sectors (see Table \ref{tab:observations} for the dates of the \TESS\ observations). For each target, we accessed the \TESS\ light curves as generated by NASA's Science Processing Operations Center (SPOC) pipeline through the Lightkurve software package \citep{Lightkurve:2018}. After receiving raw data from the spacecraft, SPOC processes the images, extracts photometry, and removes systematic errors \citep{Jenkins:2016}.  Specifically, the pipeline performs pixel-level calibrations, identifies an optimal photometric aperture and extracts the light curve, and estimates and corrects for flux contamination from nearby stars. Using the Presearch Data Conditioning (PDC) module, instrumental artifacts are removed \citep{Smith:2012, Stumpe:2012, Stumpe:2014}. The final light curves are searched for transit crossing events (TCEs) using the SPOC Transiting Planet Search (TPS, \citealt{Jenkins:2002}). The \ktwo\ targets we analyzed in this work were assigned a \TESS\ Object of Interest (TOI) number as part of the TOI catalog (N. Guerrero et al. \textit{submitted}): K2-114 b = TOI 514.01, K2-167 b = TOI 1407.01, K2-237 b = TOI 1049.01, and K2-261 b = TOI 685.01. After downloading the SPOC light curve files, we removed astrophysical variability using the Lightkurve \textit{flatten} function, which removes low frequency trends using SciPy's Savitzky-Golay filter \citep{Savitzky:1964,Lightkurve:2018,Virtanen:2020}. {In our experience removing lower frequency trends, we find no significant difference between this and a spline filter.} The \TESS\ lightcurve of K2-237 also revealed a short-period stellar variability with a period of 0.53 days, which we attributed to the nearby RR Lyrae stars (see Section \ref{sec:Variability}). We removed this short-period stellar variability by dividing out the phase-folded \TESS\ light curve at the measured variability period. {The final TESS phase-folded transits are shown in blue in Figure \ref{fig:transits}.} We use these results for fitting each system in Section \ref{sec:GlobalModel}.

\subsection{Stellar Variability in K2-237} 
\label{sec:Variability}

To search for periodic photometric variability, we analyzed the unflattened PDC version of each \ktwo\ and \tess\ light curve. First, we divided out the best-fit low-order polynomials, which effectively removed the flux trends, and retained any higher-frequency variability. The \ktwo\ light curve of K2-237 revealed a distinct M-shaped periodic modulation signal with an amplitude of around 3500 ppm. Both, the Lomb--Scargle and autocorrelation analysis indicated a period of 5.1$\pm$0.5 days during the first part of the Campaign 11 light curve. The modulation in the second part is less coherent. This is likely rotational modulation, and it being less coherent in the second part is likely due to starspot evolution, and therefore the measured rotational period of 4.7 days is less accurate but still consistent with the measurement from the first part. {This is consistent with previous analyses of the K2 observations of K2-237, which also found a 5.1-day signal of stellar rotation \citep{Soto:2018,Smith:2019}.} We removed the modulation prior to including the data set in the global fit (see Section \ref{sec:k2}).

\begin{figure}[ht!]
	\centering
	\includegraphics[width=\linewidth]{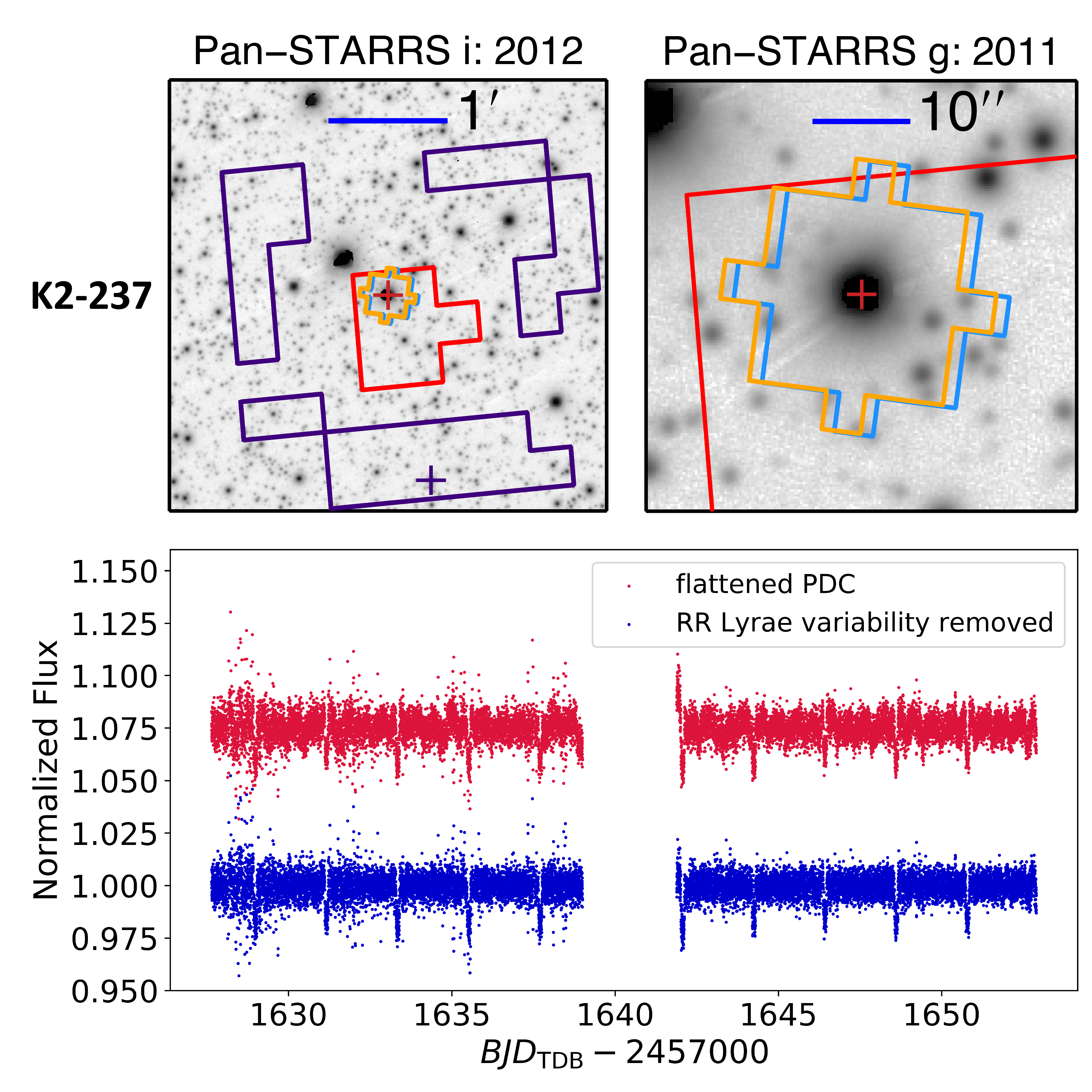}
	\caption{Pan-STARRS images of the field around K2-237 are shown in the upper panel \citep{Flewelling:2016}, including the SPOC aperture (red outline), the SPOC background apertures (purple outline), the position of the blended RR Lyrae (purple cross), and the two parts of K2 Campaign 11 apertures (blue and orange outlines). The lower panel shows the RR Lyrae signal in the TESS light curve: the flattened SPOC PDC light curve is shown before (red) and after (blue) removal of the 0.53-day variability signal.}
	\label{fig:rrlyr} 
\end{figure}

The 5-day rotational modulation signal was not detectable in the \TESS\ lightcurve of K2-237. This may be due to the fact that the \TESS\ and \ktwo\ baselines do not temporally overlap and are separated by nearly 900 days. Starspot evolution can lead to changes in observed period and amplitude, and spot constrasts in the redder TESS bandpass can also suppress amplitudes \citep{Oelkers:2018}. Instead, the \TESS\ light curve, {shown in full in Figure \ref{fig:rrlyr}}, revealed an RR-Lyrae-like signal with a period of $0.527\pm0.004$ days and an amplitude of 4200 ppm. This signal was not seen in the \ktwo\ light curve. Given that the \TESS\ pixel scale of 21 arcsec is significantly larger than the \ktwo\ pixel scale of 4 arcsec, we attributed the 0.53-day variability signal to one of the nearby contaminating stars, which was confirmed by the SPOC Data Validation \citep{Twicken:2018, Li:2019}. Specifically, in addition to detecting the signature of TOI 1049.01 (K2-237 b), a second TCE was generated with a period of 0.529 days in the SPOC Data Validation component. The difference image for this TCE in the SPOC data validation report showed a single pixel at the upper edge of the postage stamp that was highly anti-correlated with the transit signature. There was only one TIC object on that pixel (TIC 16288004). This indicates that the 0.527-day signature was introduced into the light curve through the background correction. Simbad indicates that the star at the coordinates of TIC 16288004 is a known RR Lyr variable, KY Oph (96$\arcsec$ from K2-237, {see Figure \ref{fig:rrlyr}}). We did not detect any periodic variability in any of the light curves of the other three stars.

\begin{figure}[ht!]
	\centering\vspace{.0in}
	\includegraphics[width=0.99\linewidth, trim={2.5cm 13cm 9cm 8cm}, clip]{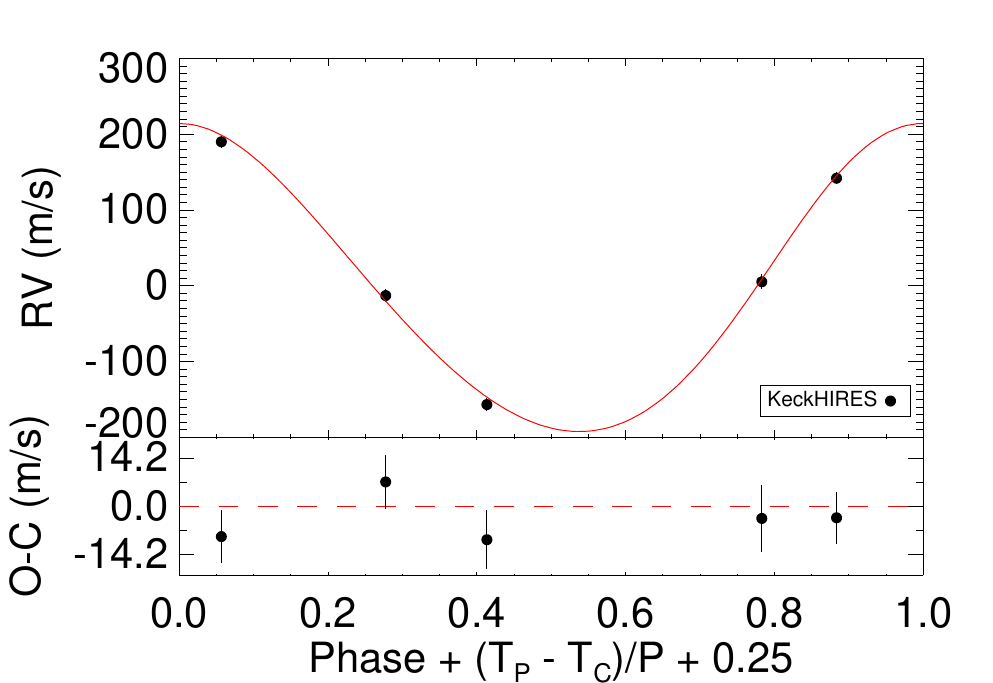}\\	\includegraphics[width=0.99\linewidth, trim={2.5cm 13cm 9cm 8cm}, clip]{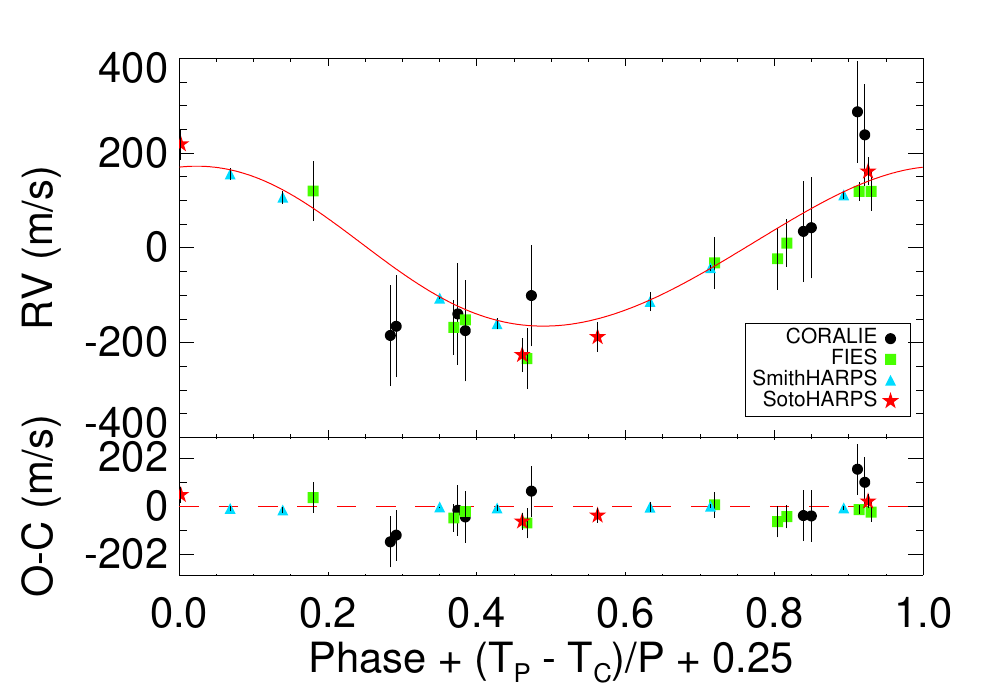}\\
	\includegraphics[width=0.99\linewidth, trim={2.5cm 13cm 9cm 8cm}, clip]{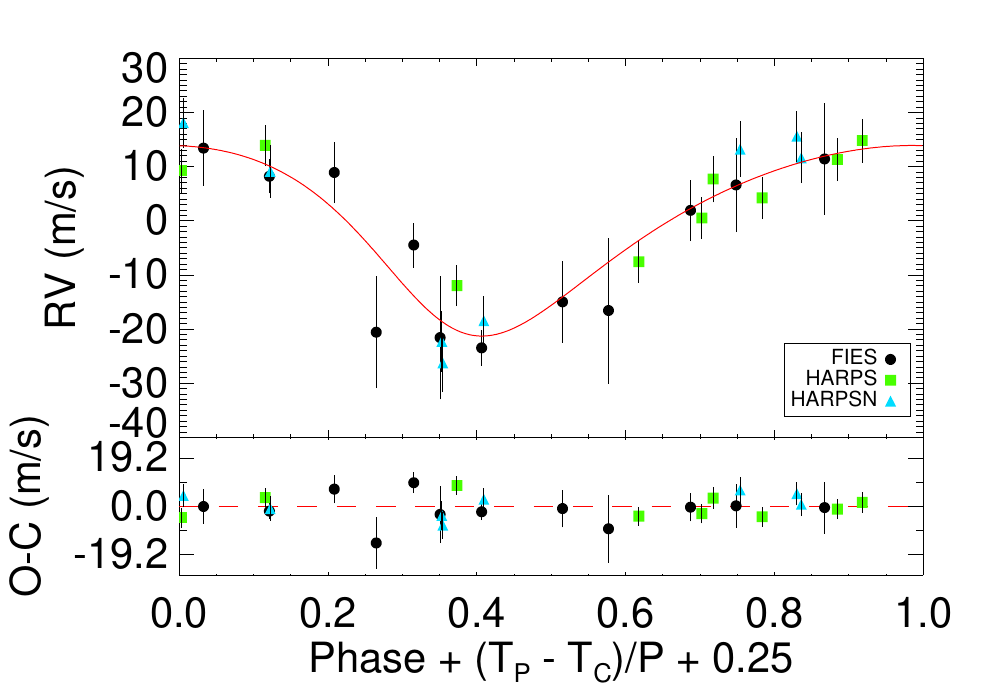}\\
	\caption{The archival radial velocity measurements for K2-114 b (top), K2-237 b (middle), and K2-261 b (bottom), phase-folded to the best fit period from our EXOFASTv2 global fit. See Section \ref{sec:RVs} for a description of the literature RVs. The EXOFASTv2 model is shown in red and the residuals to the best-fit are shown below each plot. {\(T_p\) is time of periastron and \(T_c\) is time of conjunction (transit).}}
	\label{fig:RVs} 
\end{figure}

\subsection{Archival Spectroscopy}
\label{sec:RVs}
For three of the four \Ktwo\ systems analyzed in this work (all but K2-167), we obtained the archival radial velocity (RV) measurements from the literature {(Figure \ref{fig:RVs}).} We combined these archival observations with the transit light curves extracted from the photometric observations from \Ktwo\ and \tess\ to provide updated system parameters and ephemerides for future follow up efforts. For K2-261 b we used 12 RV observations from the  FIbre-fed \'Echelle Spectrograph (FIES; \citealp{Telting:2014}), 9 RVs from the High Accuracy Radial velocity Planet Searcher for the Northern hemisphere (HARPS-N; \citealp{Cosentino:2012}), and 11 RVs from the High Accuracy Radial velocity Planet Searcher (HARPS; \citealp{Mayor:2003}) that were used in the discovery paper \citep{Johnson:2018}. For K2-114, we used the 5 RV observations from the High Resolution Echelle Spectrometer (HIRES, \citealp{Vogt:1994}) on the Keck I telescope \citep{Shporer:2017}. {With only 5 observations, the global fit has fewer degrees of freedom, and therefore a limited ability to constrain the jitter within the fit. Therefore, we provide a conservative uniform bound on the jitter variance of the fit of 300 \((m/s)^2\) on the KECK HIRES RVs for K2-114 b.} The discovery of K2-237 b was led by two separate teams \citep{Soto:2018, Smith:2019}, with RV observations coming from three separate facilities: FIES (10), HARPS (11), and the CORALIE (9) spectrograph \citep{Queloz:2000}. The 11 RV observations from HARPS were treated separately, with 4 observations coming from the \citet{Soto:2018} reduction and 7 coming from the reduction done by \citet{Smith:2019}. These observations were reduced in different manners, so we treated them as separate facilities in our global model (with unique jitter and gamma parameters). 

K2-167 b was statistically validated as part of a larger effort for \Ktwo\ campaigns 0 to 10 \citep{Mayo:2018}. There was no mass measured for this system; we did not have any RVs to include in our fit. We also used the determined metallicity from the discovery papers as a prior on our global fit (see Section \ref{sec:GlobalModel}). Specifically, we use an [Fe/H] metallicity of 0.410$\pm$0.037 dex (K2-114, \citealp{Shporer:2017}), 0.45$\pm$0.08 dex (K2-167, \citealp{Mayo:2018}), 0.14$\pm$0.05 dex (K2-237, \citealp{Soto:2018}), and 0.36$\pm$0.06 dex (K2-261, \citealp{Johnson:2018}). 

\begin{table*}
\tiny
\centering
\caption{Median values and 68\% confidence interval for global models}
\begin{tabular}{llcccccccc}
  \hline
  \hline
Parameter & Description (Units)  & K2-114 & K2-167 & K2-237 & \multicolumn{2}{c}{K2-261$^{\star}$}\\
  \hline
Probability\dotfill & & 100 \%& 100 \%& 100 \%& 75.7 \% & 24.2 \% \\
\multicolumn{2}{l}{Stellar Parameters:}&\\
~~~~$M_*$\dotfill &Mass (\msun)\dotfill &$0.860^{+0.039}_{-0.034}$&$1.337\pm0.069$&$1.263^{+0.052}_{-0.068}$&$1.090^{+0.050}_{-0.051}$&$1.266^{+0.043}_{-0.041}$\\
~~~~$R_*$\dotfill &Radius (\rsun)\dotfill &$0.824^{+0.024}_{-0.023}$&$1.460\pm0.051$&$1.261^{+0.031}_{-0.029}$&$1.642^{+0.058}_{-0.059}$&$1.641^{+0.058}_{-0.055}$\\
~~~~$L_*$\dotfill &Luminosity (\lsun)\dotfill &$0.359^{+0.032}_{-0.030}$&$2.73^{+0.26}_{-0.24}$&$2.35^{+0.30}_{-0.27}$&$2.13^{+0.20}_{-0.19}$&$2.22^{+0.21}_{-0.19}$\\
~~~~$\rho_*$\dotfill &Density (cgs)\dotfill &$2.17^{+0.18}_{-0.17}$&$0.606^{+0.068}_{-0.061}$&$0.888^{+0.065}_{-0.075}$&$0.346^{+0.038}_{-0.032}$&$0.405^{+0.038}_{-0.035}$\\
~~~~$\log{g}$\dotfill &Surface gravity (cgs)\dotfill &$4.542\pm0.025$&$4.236\pm0.034$&$4.339^{+0.022}_{-0.031}$&$4.044^{+0.031}_{-0.030}$&$4.111\pm0.025$\\
~~~~$T_{\rm eff}$\dotfill &Effective Temperature (K)\dotfill &$4920^{+68}_{-69}$&$6141^{+92}_{-93}$&$6360^{+190}_{-200}$&$5445^{+77}_{-76}$&$5503^{+73}_{-74}$\\
~~~~$[{\rm Fe/H}]$\dotfill &Metallicity (dex)\dotfill &$0.410^{+0.036}_{-0.037}$&$0.425^{+0.058}_{-0.070}$&$0.137^{+0.050}_{-0.049}$&$0.360\pm0.059$&$0.384^{+0.057}_{-0.058}$\\
~~~~$[{\rm Fe/H}]_{0}$\dotfill &Initial Metallicity \dotfill &$0.378\pm0.046$&$0.433^{+0.044}_{-0.060}$&$0.152^{+0.050}_{-0.055}$&$0.359^{+0.057}_{-0.059}$&$0.382^{+0.053}_{-0.055}$\\
~~~~$Age$\dotfill &Age (Gyr)\dotfill &$7.2^{+4.3}_{-4.5}$&$2.1^{+1.4}_{-1.1}$&$1.02^{+1.6}_{-0.74}$&$9.3^{+2.0}_{-1.6}$&$4.83^{+0.68}_{-0.71}$\\
~~~~$EEP$\dotfill &Equal Evolutionary Phase \dotfill &$345^{+21}_{-28}$&$346^{+32}_{-21}$&$324^{+29}_{-40}$&$455.7^{+3.7}_{-5.1}$&$415.4^{+8.0}_{-12}$\\
~~~~$A_V$\dotfill &V-band extinction (mag)\dotfill &---&---&$0.31^{+0.12}_{-0.13}$&---&---\\
~~~~$\sigma_{SED}$\dotfill &SED photometry error scaling \dotfill &---&---&$1.72^{+0.72}_{-0.43}$&---&---\\
\smallskip\\\multicolumn{2}{l}{Planetary Parameters:}&b\smallskip\\
~~~~$P$\dotfill &Period (days)\dotfill &$11.3909311\pm0.0000034$&$9.978570\pm0.000022$&$2.18053539^{+0.00000086}_{-0.00000085}$&$11.633478\pm0.000017$&$11.633478\pm0.000017$\\
~~~~$R_P$\dotfill &Radius (\rj)\dotfill &$0.932\pm0.031$&$0.202^{+0.014}_{-0.010}$&$1.445^{+0.049}_{-0.045}$&$0.850^{+0.035}_{-0.034}$&$0.848^{+0.035}_{-0.032}$\\
~~~~$M_P$\dotfill &Mass (\mj)\dotfill &$2.01\pm0.12$&---&$1.363^{+0.11}_{-0.092}$&$0.188\pm0.025$&$0.213^{+0.026}_{-0.028}$\\
~~~~$T_C$\dotfill &Time of conjunction (\bjdtdb)\dotfill &$2457140.32399\pm0.00023$&$2456979.9326\pm0.0020$&$2457656.463880^{+0.000037}_{-0.000036}$&$2457906.84105^{+0.00030}_{-0.00035}$&$2457906.84103^{+0.00029}_{-0.00032}$\\
~~~~$T_0^\dagger$\dotfill &Optimal conjunction Time (\bjdtdb)\dotfill &$2457664.30683^{+0.00016}_{-0.00017}$&$2457349.1397\pm0.0018$&$2457702.255123^{+0.000032}_{-0.000031}$&$2457976.64192^{+0.00028}_{-0.00033}$&$2457976.64189^{+0.00027}_{-0.00031}$\\
~~~~$a$\dotfill &Semi-major axis (AU)\dotfill &$0.0943^{+0.0014}_{-0.0012}$&$0.0999^{+0.0017}_{-0.0018}$&$0.03558^{+0.00048}_{-0.00065}$&$0.1034^{+0.0015}_{-0.0016}$&$0.1087\pm0.0012$\\
~~~~$i$\dotfill &Inclination (Degrees)\dotfill &$89.21^{+0.22}_{-0.14}$&$87.41^{+1.5}_{-0.62}$&$88.03^{+1.1}_{-0.82}$&$88.07^{+1.1}_{-0.65}$&$88.34^{+1.0}_{-0.62}$\\
~~~~$e$\dotfill &Eccentricity \dotfill &$0.081^{+0.031}_{-0.030}$&$0.41^{+0.31}_{-0.22}$&$0.042^{+0.034}_{-0.028}$&$0.286^{+0.064}_{-0.069}$&$0.248^{+0.065}_{-0.063}$\\
~~~~$\omega_*$\dotfill &Argument of Periastron (Degrees)\dotfill &$-51^{+21}_{-13}$&$20^{+110}_{-170}$&$74\pm38$&$131^{+16}_{-17}$&$137\pm17$\\
~~~~$T_{eq}$\dotfill &Equilibrium temperature (K)\dotfill &$701\pm14$&$1131\pm23$&$1828^{+48}_{-46}$&$1046^{+21}_{-22}$&$1030\pm21$\\
~~~~$K$\dotfill &RV semi-amplitude (m/s)\dotfill &$200\pm11$&---&$183^{+15}_{-11}$&$16.7\pm2.2$&$16.9^{+2.1}_{-2.2}$\\
~~~~$\log{K}$\dotfill &Log of RV semi-amplitude \dotfill &$2.303^{+0.022}_{-0.024}$&---&$2.263^{+0.035}_{-0.028}$&$1.223^{+0.054}_{-0.063}$&$1.227^{+0.051}_{-0.061}$\\
~~~~$R_P/R_*$\dotfill &Radius of planet in stellar radii \dotfill &$0.1163^{+0.0014}_{-0.0016}$&$0.01417^{+0.00086}_{-0.00047}$&$0.1177^{+0.0027}_{-0.0026}$&$0.05304^{+0.0011}_{-0.00074}$&$0.05295^{+0.0011}_{-0.00068}$\\
~~~~$a/R_*$\dotfill &Semi-major axis in stellar radii \dotfill &$24.63^{+0.67}_{-0.65}$&$14.72^{+0.53}_{-0.52}$&$6.07^{+0.14}_{-0.18}$&$13.53^{+0.48}_{-0.44}$&$14.26^{+0.43}_{-0.42}$\\
~~~~$\delta$\dotfill &Transit depth (fraction)\dotfill &$0.01352^{+0.00033}_{-0.00036}$&$0.000201^{+0.000025}_{-0.000013}$&$0.01385^{+0.00065}_{-0.00060}$&$0.002814^{+0.00012}_{-0.000078}$&$0.002804^{+0.00012}_{-0.000072}$\\
~~~~$\tau$\dotfill &Ingress/egress transit duration (days)\dotfill &$0.0195\pm0.0015$&$0.00271^{+0.0022}_{-0.00062}$&$0.01348^{+0.00056}_{-0.00047}$&$0.0122^{+0.0023}_{-0.0014}$&$0.0120^{+0.0023}_{-0.0013}$\\
~~~~$T_{14}$\dotfill &Total transit duration (days)\dotfill &$0.1654\pm0.0012$&$0.1513^{+0.0039}_{-0.0038}$&$0.12199^{+0.00039}_{-0.00037}$&$0.2141^{+0.0020}_{-0.0014}$&$0.2141^{+0.0019}_{-0.0014}$\\
~~~~$T_{FWHM}$\dotfill &FWHM transit duration (days)\dotfill &$0.14583^{+0.00086}_{-0.00085}$&$0.1478^{+0.0034}_{-0.0036}$&$0.10849^{+0.00026}_{-0.00028}$&$0.20176^{+0.00085}_{-0.00084}$&$0.20182^{+0.00085}_{-0.00083}$\\
~~~~$b$\dotfill &Transit Impact parameter \dotfill &$0.361^{+0.067}_{-0.10}$&$0.48^{+0.26}_{-0.32}$&$0.201^{+0.085}_{-0.12}$&$0.35^{+0.15}_{-0.21}$&$0.33^{+0.16}_{-0.21}$\\
~~~~$b_S$\dotfill &Eclipse impact parameter \dotfill &$0.317^{+0.047}_{-0.079}$&$0.51^{+0.23}_{-0.32}$&$0.216^{+0.084}_{-0.12}$&$0.53^{+0.14}_{-0.30}$&$0.47^{+0.14}_{-0.28}$\\
~~~~$\tau_S$\dotfill &Ingress/egress eclipse duration (days)\dotfill &$0.01691^{+0.0011}_{-0.00098}$&$0.0039^{+0.0014}_{-0.0019}$&$0.01450^{+0.0010}_{-0.00077}$&$0.0206^{+0.0033}_{-0.0024}$&$0.0180^{+0.0021}_{-0.0017}$\\
~~~~$T_{S,14}$\dotfill &Total eclipse duration (days)\dotfill &$0.1481^{+0.0090}_{-0.0079}$&$0.156^{+0.084}_{-0.036}$&$0.1303^{+0.0087}_{-0.0069}$&$0.298^{+0.058}_{-0.046}$&$0.284^{+0.043}_{-0.039}$\\
~~~~$T_{S,FWHM}$\dotfill &FWHM eclipse duration (days)\dotfill &$0.1311^{+0.0085}_{-0.0071}$&$0.152^{+0.085}_{-0.035}$&$0.1158^{+0.0079}_{-0.0063}$&$0.278^{+0.059}_{-0.048}$&$0.266^{+0.044}_{-0.040}$\\
~~~~$\delta_{S,3.6\mu m}$\dotfill &Blackbody eclipse depth at 3.6$\mu$m (ppm)\dotfill &$53.4^{+6.0}_{-5.5}$&$5.42^{+0.76}_{-0.53}$&$1507^{+85}_{-79}$&$66.3^{+5.6}_{-5.3}$&$61.3^{+5.3}_{-4.6}$\\
~~~~$\delta_{S,4.5\mu m}$\dotfill &Blackbody eclipse depth at 4.5$\mu$m (ppm)\dotfill &$129^{+12}_{-11}$&$8.68^{+1.2}_{-0.77}$&$1900^{+100}_{-93}$&$111.2^{+7.9}_{-7.4}$&$104.1^{+7.6}_{-6.6}$\\
~~~~$\rho_P$\dotfill &Density (cgs)\dotfill &$3.08^{+0.37}_{-0.33}$&---&$0.562^{+0.066}_{-0.061}$&$0.380^{+0.071}_{-0.063}$&$0.430^{+0.074}_{-0.068}$\\
~~~~$logg_P$\dotfill &Surface gravity \dotfill &$3.758\pm0.038$&---&$3.210^{+0.040}_{-0.039}$&$2.810^{+0.063}_{-0.070}$&$2.864^{+0.059}_{-0.066}$\\
~~~~$\fave$\dotfill &Incident Flux (\fluxcgs)\dotfill &$0.0544^{+0.0045}_{-0.0042}$&$0.312^{+0.049}_{-0.078}$&$2.53^{+0.28}_{-0.25}$&$0.250^{+0.021}_{-0.020}$&$0.240^{+0.020}_{-0.018}$\\
~~~~$T_P$\dotfill &Time of Periastron (\bjdtdb)\dotfill &$2457136.05^{+0.66}_{-0.43}$&$2456979.86^{+0.87}_{-0.86}$&$2457656.38\pm0.21$&$2457895.93^{+0.34}_{-0.28}$&$2457896.10^{+0.41}_{-0.32}$\\
~~~~$T_S$\dotfill &Time of eclipse (\bjdtdb)\dotfill &$2457146.38^{+0.11}_{-0.12}$&$2456974.9^{+2.9}_{-3.2}$&$2457657.565^{+0.034}_{-0.025}$&$2457899.69^{+0.55}_{-0.64}$&$2457899.75^{+0.48}_{-0.57}$\\
~~~~$T_A$\dotfill &Time of Ascending Node (\bjdtdb)\dotfill &$2457137.45\pm0.11$&$2456978.1^{+1.2}_{-1.5}$&$2457658.129^{+0.031}_{-0.027}$&$2457904.24^{+0.35}_{-0.39}$&$2457904.06^{+0.33}_{-0.39}$\\ 
    ~~~~$T_D$\dotfill &Time of Descending Node (\bjdtdb)\dotfill &$2457143.59^{+0.16}_{-0.18}$&$2456981.8^{+1.5}_{-1.2}$&$2457656.993^{+0.019}_{-0.026}$&$2457896.86^{+0.28}_{-0.25}$&$2457897.00^{+0.27}_{-0.26}$\\
~~~~$e\cos{\omega_*}$\dotfill & \dotfill &$0.049^{+0.014}_{-0.017}$&$0.00\pm0.49$&$0.008^{+0.025}_{-0.018}$&$-0.177^{+0.073}_{-0.087}$&$-0.170^{+0.065}_{-0.078}$\\
~~~~$e\sin{\omega_*}$\dotfill & \dotfill &$-0.062^{+0.036}_{-0.034}$&$0.07^{+0.21}_{-0.30}$&$0.034^{+0.033}_{-0.029}$&$0.211^{+0.058}_{-0.071}$&$0.168^{+0.054}_{-0.070}$\\
~~~~$M_P\sin i$\dotfill &Minimum mass (\mj)\dotfill &$2.01\pm0.12$&---&$1.362^{+0.11}_{-0.092}$&$0.188\pm0.025$&$0.212^{+0.026}_{-0.028}$\\
~~~~$M_P/M_*$\dotfill &Mass ratio \dotfill &$0.00222\pm0.00012$&---&$0.001034^{+0.000090}_{-0.000066}$&$0.000165^{+0.000021}_{-0.000022}$&$0.000160^{+0.000020}_{-0.000021}$\\
\hline\\
\end{tabular}
 \begin{flushleft} 
  \footnotesize{ 
    \textbf{NOTES:}\\$^\star$The global solution for K2-261 b showed a clear bimodality in the host star's mass and age (see Figure \ref{fig:PDF} and \S\ref{sec:bimodality}). We extract a solution and a probability for each peak, which are both shown in the table. The lower stellar mass (and high age) solution is significantly more likely, but both solutions are presented for future studies on K2-261 b.\\
See Table 3 in \citet{Eastman:2019} for a list of the derived and fitted parameters in EXOFASTv2.\\
$^\dagger$Minimum covariance with period.\\
All values in this table for the secondary occultation are predicted values from our global analysis. \\
See \S\ref{sec:GlobalModel} for a description of how the EXOFASTv2 fit was conducted and what priors were used for each fit.\\ 
               }
 \end{flushleft}
\label{tab:exofast_planetary}
\end{table*}

\begin{table*}
\scriptsize
\centering
\caption{Median values and 68\% confidence interval for global model}
\begin{tabular}{llcccccc}
  \hline
  \hline
\smallskip\\\multicolumn{2}{l}{Wavelength Parameters:}&Kepler&TESS\smallskip\\
K2-114&&\\
~~~~$u_{1}$\dotfill &linear limb-darkening coeff \dotfill &$0.612^{+0.026}_{-0.027}$&$0.466\pm0.050$\\
~~~~$u_{2}$\dotfill &quadratic limb-darkening coeff \dotfill &$0.116\pm0.036$&$0.189\pm0.050$\\
~~~~$A_D$\dotfill &Dilution from neighboring stars \dotfill &--&$-0.085\pm0.060$\\
K2-167&&\\
~~~~$u_{1}$\dotfill &linear limb-darkening coeff \dotfill &$0.381\pm0.051$&$0.260\pm0.051$\\
~~~~$u_{2}$\dotfill &quadratic limb-darkening coeff \dotfill &$0.306\pm0.050$&$0.302\pm0.050$\\
~~~~$A_D$\dotfill &Dilution from neighboring stars \dotfill &--&$-0.14\pm0.12$\\
K2-237&&\\
~~~~$u_{1}$\dotfill &linear limb-darkening coeff \dotfill &$0.329^{+0.014}_{-0.015}$&$0.226\pm0.042$\\
~~~~$u_{2}$\dotfill &quadratic limb-darkening coeff \dotfill &$0.266^{+0.033}_{-0.034}$&$0.310\pm0.048$\\
~~~~$A_D$\dotfill &Dilution from neighboring stars \dotfill &$-0.017\pm0.047$&$-0.097^{+0.053}_{-0.054}$\\
K2-261&&\\
~~~~$u_{1}$\dotfill &linear limb-darkening coeff \dotfill &$0.481^{+0.030}_{-0.031}$&$0.376\pm0.044$\\
~~~~$u_{2}$\dotfill &quadratic limb-darkening coeff \dotfill &$0.193\pm0.046$&$0.250\pm0.048$\\
~~~~$A_D$\dotfill &Dilution from neighboring stars \dotfill &--&$-0.046^{+0.025}_{-0.026}$\\
\hline
\smallskip\\K2-114&&\\
\multicolumn{2}{l}{Velocity Parameters:}&Keck HIRES\smallskip\\
~~~~$\gamma_{\rm rel}$\dotfill &Relative RV Offset (m/s)\dotfill &$-41.0^{+6.7}_{-6.8}$\\
~~~~$\sigma_J$\dotfill &RV Jitter (m/s)\dotfill &$11.9^{+3.8}_{-4.9}$\\
~~~~$\sigma_J^2$\dotfill &RV Jitter Variance \dotfill &$141^{+100}_{-93}$\\
\multicolumn{2}{l}{Transit Parameters:}&K2 C5 (Kepler)&K2 C18 (Kepler)&TESS\smallskip\\
~~~~$\sigma^{2}$\dotfill &Added Variance \dotfill &$0.000000012^{+0.000000018}_{-0.000000016}$&$0.00000082\pm0.00000013$&$-0.0002645^{+0.0000054}_{-0.0000050}$\\
~~~~$F_0$\dotfill &Baseline flux \dotfill &$1.000034\pm0.000032$&$0.999746^{+0.000043}_{-0.000044}$&$1.00005^{+0.00040}_{-0.00039}$\\
K2-167&&\\
\multicolumn{2}{l}{Transit Parameters:}& &K2 C3 (Kepler)&TESS\smallskip\\
~~~~$\sigma^{2}$\dotfill &Added Variance \dotfill &$-0.00000000008^{+0.00000000021}_{-0.00000000018}$&$-0.0000000020^{+0.0000000049}_{-0.0000000046}$\\
~~~~$F_0$\dotfill &Baseline flux \dotfill &$0.9999988\pm0.0000042$&$1.000017\pm0.000013$\\
K2-237&&\\
\multicolumn{2}{l}{Velocity Parameters:}&CORALIE&FIES&HARPS (Smith)&HARPS (Soto)\smallskip\\
~~~~$\gamma_{\rm rel}$\dotfill &Relative RV Offset (m/s)\dotfill &$-22250^{+40}_{-41}$&$-22501^{+16}_{-17}$&$-22325.5^{+8.8}_{-9.5}$&$-22252\pm14$\\
~~~~$\sigma_J$\dotfill &RV Jitter (m/s)\dotfill &$110^{+49}_{-30}$&$0.00^{+42}_{-0.00}$&$18^{+23}_{-18}$&$6.6^{+8.3}_{-6.6}$\\
~~~~$\sigma_J^2$\dotfill &RV Jitter Variance \dotfill &$12200^{+13000}_{-5800}$&$-30^{+1800}_{-560}$&$330^{+1400}_{-330}$&$40^{+180}_{-220}$\\
\multicolumn{2}{l}{Transit Parameters:}&K2 (Kepler)&K2 (Kepler)&TESS\smallskip\\
~~~~$\sigma^{2}$\dotfill &Added Variance \dotfill &$0.0000000056^{+0.0000000021}_{-0.0000000018}$&$0.00000000388^{+0.0000000010}_{-0.00000000094}$&$0.00001815^{+0.00000079}_{-0.00000076}$\\
~~~~$F_0$\dotfill &Baseline flux \dotfill &$0.9999999\pm0.0000097$&$0.9999946\pm0.0000062$&$1.00009\pm0.00011$\\
K2-261&&\\
\multicolumn{2}{l}{Velocity Parameters:}&FIES&HARPS&HARPSN\smallskip\\
~~~~$\gamma_{\rm rel}$\dotfill &Relative RV Offset (m/s)\dotfill &$-13.8^{+2.6}_{-2.7}$&$3341.6^{+1.8}_{-2.0}$&$3334.9^{+2.5}_{-3.\
0}$\\
~~~~$\sigma_J$\dotfill &RV Jitter (m/s)\dotfill &$4.2^{+3.9}_{-4.2}$&$5.0^{+2.6}_{-1.7}$&$6.8^{+4.6}_{-2.9}$\\
~~~~$\sigma_J^2$\dotfill &RV Jitter Variance \dotfill &$17^{+47}_{-22}$&$24^{+32}_{-14}$&$46^{+85}_{-31}$\\
~~~~$\sigma^{2}$\dotfill &Added Variance \dotfill &$0.00000000157^{+0.00000000064}_{-0.00000000057}$&$-0.000000194^{+0.000000042}_{-0.000000041}$\\
~~~~$F_0$\dotfill &Baseline flux \dotfill &$0.9999979\pm0.0000060$&$1.000031\pm0.000033$\\
\multicolumn{2}{l}{Transit Parameters:}&K2 (Kepler)&TESS \smallskip\\
\hline
  \hline
\label{tab:exofast_other}
 \end{tabular}
\begin{flushleft}
\textbf{NOTES:}\\
$^\star$The global solution for K2-261 b showed a clear bimodality in the host star's mass and age (see Figure \ref{fig:PDF} and Section \ref{sec:bimodality}). The transit, velocity, and wavelength parameters shown in this table for K2-261 are for the preferred solution only.\\
$^{\star\star}$The RV jitter variance for K2-114 b was constrained to 300 \((m/s)^2\).\\
See Table 3 in \citet{Eastman:2019} for a list of the derived and fitted parameters in EXOFASTv2.\\
All values in this table for the secondary occultation are predicted values from our global analysis. \\
See \S\ref{sec:GlobalModel} for a description of how the EXOFASTv2 fit was conducted and what priors were used for each fit. 
  \end{flushleft}
\end{table*}

\section{EXOFAST\lowercase{v}2 Global Fits} 
\label{sec:GlobalModel}
The advent of new software packages allows us to more easily combine the archival observations (\Ktwo\ photometry and RVs) with new data from \tess. Additionally, with the ongoing success of the {\it Gaia} mission, we now know the distances to almost every known planet host, allowing us to accurately characterize the host star through a combination of spectral energy distributions, {\it Gaia} parallaxes, and updated stellar models. To refine the ephemerides and system parameters of the four \Ktwo\ systems, we used the exoplanet fitting suite, EXOFASTv2 \citep{Eastman:2013, Eastman:2017, Eastman:2019}. In each case other than K2-237 (see Section \ref{sec:K2-237}), we first conducted a preliminary fit of the entire system using EXOFASTv2 to get an estimate for the surface gravity of the host star. We then performed a fit of the spectral energy distribution (SED) of the host star with EXOFASTv2, using the determined {stellar surface gravity (\logg)} as a starting point with a loose 0.25 dex {Gaussian prior.} We also included Gaussian priors on the {stellar metallicity (\feh) from the discovery paper (see Section \ref{sec:RVs})} and parallax from {\it Gaia}, and constrained the maximum line of sight extinction (A$_V$) for each system using the galactic dust maps from \citet{Schlegel:1998} and \citet{Schlafly:2011}. The precision on fundamental stellar parameters like {\feh\ and the stellar effective temperature (\teff)} should be limited to the precision of stellar radii measurements from interferometry \citep{White:2018}. The SED fit provided values on the stellar effective temperature (\teff\ ) and stellar radius (\rstar) that were too precise, so we used the resulting \teff\ and \rstar\ with the adopted fractional errors of 1.5\% (for \teff) and 3.5\% (for \rstar\ ) as Gaussian priors on the full global fit. This resulted in a prior on \rstar\ and \teff\ of 0.810$\pm$0.0284 \rsun\ and 4930.0$\pm$82.0 K for K2-114, 1.664$\pm$0.058 \rsun\ and 6182.0$\pm$93.0 K for K2-167, and 1.467$\pm$0.052 \rsun\ and 5449.0$\pm$82.0 K for K2-261. Specifically, for each system we simultaneously fit the \tess\ (see Section \ref{sec:TESS}) and \Ktwo\ (see Section \ref{sec:k2}) photometry with the archival RV data (see Section \ref{sec:RVs}). In the case of K2-167, no radial velocity observations are available \citep{Mayo:2018}, so we globally modeled only the photometric data. Within the fit, the mass, radius, and age of the host star are are constrained by the MESA Isochrones and Stellar Tracks (MIST) stellar evolution models \citep{Dotter:2016, Choi:2016, Paxton:2011, Paxton:2013, Paxton:2015}. {In every case, we allow EXOFASTv2 to fit a dilution term to the \tess\ light curve, using the the \Ktwo\ lightcurve as a reference. In all cases, the the dilution term on the \tess\ light curve is consistent with zero at $<$2$\sigma$. Since we adjusted the \Ktwo\ light curve from the standard pipeline (see Section \ref{sec:k2}), we allow EXOFASTv2 to fit a dilution term for the \ktwo\ lightcurve as well in this fit but bound it with a conservative 5\% prior around zero. The dilution term for the \Ktwo\ lightcurve is consistent with zero at $<$1$\sigma$. In all cases, we aimed for strict convergence criteria that required a Gelman--Rubin statistic \citep{Gelman:1992}} of less than 1.01 and at least 1000 independent draws in each parameter. The results are shown in Figures \ref{fig:transits} and \ref{fig:RVs}, and Tables \ref{tab:exofast_planetary} and  \ref{tab:exofast_other}. 

\begin{figure*}[!ht]
	\centering\vspace{.0in}
	\includegraphics[width=0.9\linewidth, trim={0cm 0cm 0cm 0cm}, clip]{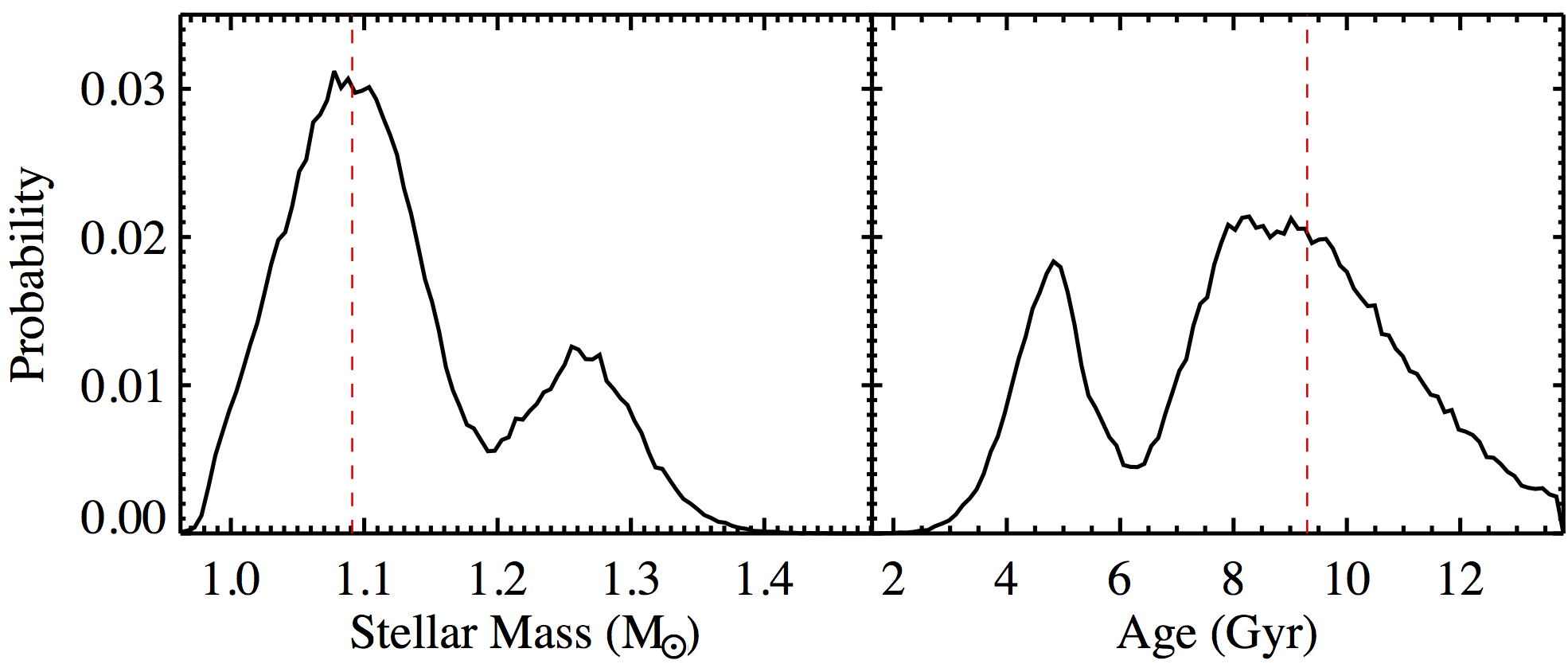}
	\caption{The probability distribution function for (Left) M$_{\rm star}$ and (Right) Age of K2-261 b from our global fit. The red line shows the median value for each parameter from the adopted solution (see Section \ref{sec:GlobalModel}).}
	\label{fig:PDF} 
\end{figure*}

\subsection{K2-237 Global Fit}
\label{sec:K2-237}
For the K2-237 system, we ran the full global model slightly differently than as described in the previous section. Specifically, the SED was far less constraining on the fundamental stellar parameters, due to the target lacking Tycho B$_T$ and V$_T$ magnitudes and the maximum line-of-sight extinction being higher than is typical. Therefore, we included the SED within the global model, placing Gaussian priors on the {\it Gaia} parallax and {the \feh\ from \citet{Soto:2018} (see Section \ref{sec:RVs})} and an upper limit on the maximum line of sight extinction (A$_V$) from \citet{Schlegel:1998} and \citet{Schlafly:2011}. The resulting parameters have larger uncertainties than for the other systems.



\subsection{K2-261 Bimodality}
\label{sec:bimodality}
When analyzing the probability distribution function (PDF) for the host star's mass and age for K2-261, we noticed a bimodality (see Figure \ref{fig:PDF}). The peaks are centered at 1.09 \msun\ (9.3 Gyr) and 1.27 \msun\ (4.84 Gyr), with the lower mass solution having a 76.1\% probability of being correct from our analysis. With no optimal way to properly represent the PDF due to the bimodality, we split the host star's mass at the minimum value between the two peaks in the posterior distribution at M$_{\star}$ = 1.19 \msun, and extracted two solutions, one for each mass peak. The two solutions are shown in Table \ref{tab:exofast_planetary}. {There are no significant differences in the systematic parameters resulting from the two solutions; we believe the bimodality is astrophysical.} We adopt the solution for M$_{\star}$ = 1.09 \msun, but we present both solutions, as they may be important for future studies on K2-261 b.

\begin{table}
\centering\scriptsize
\caption{The discovery ephemerides (\citet{Livingston:2019}, \citet{Mayo:2018}, \citet{Smith:2019}, \citet{Johnson:2018}) and our updated ephemerides, and the 3-sigma uncertainty on the predicted transit times for the years 2020, 2025, and 2030.}
\begin{tabular}{r|l|l}
\multicolumn{1}{l}{} & Discovery & Updated \\
\hline
\hline
\multicolumn{3}{l}{\textbf{K2-114}} \\
P$^{\star}$ & (11.391013$^{+0.000224}_{-0.000225}$) d & (11.3909311 $\pm$ 0.0000034) d \\
T$_0$ & (2457151.71493$\pm$0.00069) BJD & (2457664.30683$^{+0.00016}_{-0.00017}$) BJD \\
3$\sigma_{2020}$ & 2.5 hours & 3 minutes \\
3$\sigma_{2025}$ & 5.1 hours & 5 minutes \\
3$\sigma_{2030}$ & 7.6 hours & 8 minutes \\
\hline
\multicolumn{3}{l}{\textbf{K2-167}} \\
P & (9.977481$^{+0.001039}_{-0.001007}$) d& (9.978570$\pm$0.000022) d \\
T$_0$ & (2456979.93678$^{+0.002518}_{-0.002443}$) BJD & (2457349.1397$\pm$0.0018) BJD \\
3$\sigma_{2020}$ & 14 hours & 26 minutes \\
3$\sigma_{2025}$ & 27.5 hours & 45 minutes \\
3$\sigma_{2030}$ & 40.6 hours & 1.1 hours \\\hline
\multicolumn{3}{l}{\textbf{K2-237}} \\
P$^{\star\star}$ & (2.1805577 $\pm$ 0.0000057) d& (2.18053539$^{+0.00000086}_{-0.00000085}$) d\\
T$_0$ & (2457656.4633789$\pm$0.0000048) BJD &(2457702.255123$^{+0.000032}_{-0.000031}$) BJD \\
3$\sigma_{2020}$ & 13 minutes & 2 minutes \\
3$\sigma_{2025}$ & 34 minutes & 5 minutes \\
3$\sigma_{2030}$ & 55 minutes & 8 minutes \\\hline
\multicolumn{3}{l}{\textbf{K2-261}} \\
P & (11.63344 $\pm$ 0.00012) d & (11.633478 $\pm$ 0.000017) d \\
T$_0$ & (2457906.84084$^{0.00054}_{-0.00067}$) BJD & (2457976.64192$^{+0.00028}_{-0.00033}$) BJD \\
3$\sigma_{2020}$ & 45 minutes & 7 minutes \\
3$\sigma_{2025}$ & 2.1 hours & 19 minutes \\
3$\sigma_{2030}$ & 3.5 hours & 30 minutes
\end{tabular}
\label{tab:comparison}
 \begin{flushleft} 
  \footnotesize{ 
    \textbf{\textsc{NOTES:}}
$^{\star}$The new period for K2-114 b is also consistent with the period reported in the earlier discovery paper \citep{Shporer:2017}.\\
$^{\star\star}$The new period for K2-237 b is $\sim4\sigma$ discrepant with \citet{Smith:2019} reported period but consistent with less precise period reported by \citet{Soto:2018}.}
 \end{flushleft}
\end{table}

\section{Discussion}
\label{sec:discussion}
With the focus of future missions (like {\it JWST}) centered on studying the atmospheres of exoplanets through transmission spectroscopy, precise ephemerides and updated parameters will be crucial for scheduling and interpreting these observations. Additionally, some of these facilities will have a fixed lifetime and very high operating costs. Therefore, we want to be as efficient as possible in using these precious resources, allowing us to maximize their scientific productivity. In this paper, we have presented a case study of four known \ktwo\ planetary systems that were observed in the first year of NASA's \tess\ mission. Our results show that combining the \ktwo\ and \TESS\ data sets (along with archival spectroscopy and {\it Gaia} parallaxes) can reduce the uncertainty on the time of future transit by up to an order of magnitude compared to the \ktwo\ discovery results. The original and updated ephemerides, with their 1-sigma confidence intervals, are found in Table \ref{tab:comparison}. As compared to the discovery ephemerides, we reduced the uncertainty on the planet's period by roughly a factor of 66 for K2-114 \citep{Livingston:2019}, a factor of 44 for K2-167 \citep{Mayo:2018}, and a factor of 7 for both K2-237 \citep{Smith:2019} and K2-261 \citep{Johnson:2018}. {For K2-237, we compare to \citet{Smith:2019} with a 4-sigma discrepancy, but our period is consistent with the other discovery ephemeris from \citet{Soto:2018}. The reason for the discrepancy in the discovery paper results is unclear. \citet{Edwards:2020} independently refined the ephemeris of K2-237 using TESS observations and found a period that is consistent to 1 sigma with our result.} Our results also provide updated planetary parameters for each system (see Table \ref{tab:exofast_planetary}), which will be important for interpreting any future follow up results.

\begin{figure}[ht!]
	\centering
	\includegraphics[width=0.8\linewidth]{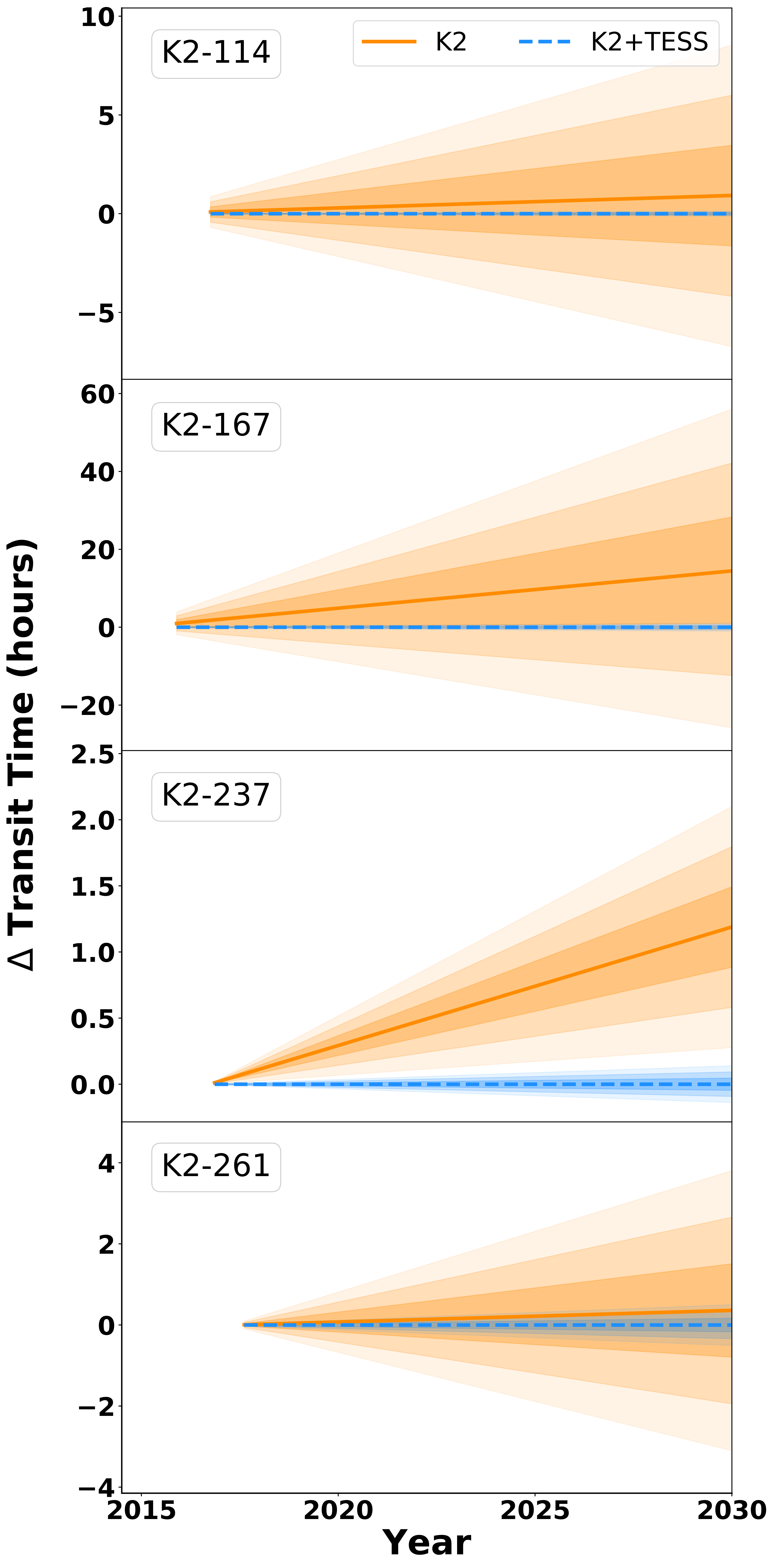}
	\caption{Difference in time of transit predicted by the \ktwo\ discovery papers as compared to the updated predictions from this work, projected to the year 2030. The shaded regions indicate 1$\sigma$, 2$\sigma$, and 3$\sigma$ confidence intervals. For K2-114 b, our ephemeris is consistent with both \citet{Livingston:2019} (shown here) and \citet{Shporer:2017} (not shown). For K2-237 b, our new ephemeris is $\sim$4$\sigma$ discrepant to the ephemeris from \citet{Smith:2019} (shown here) but is consistent with less precise period reported by \citet{Soto:2018} (not shown).}
	\label{fig:JWST} 
\end{figure}

Prior to our analysis, the uncertainties on the periods from the \ktwo\ discovery papers propagate to produce high uncertainty (hours to days) on the predicted transit times within the next decade (see Figure \ref{fig:JWST}). The 3-sigma uncertainties on the transit times predicted by the discovery ephemerides, shown in Table \ref{tab:comparison}, are as high as 40.6 hours by 2030 (for K2-167). Because the addition of the \TESS\ data enables such major improvement in precision on the period, the uncertainties on our predicted transit times are dominated by the uncertainty on the time of conjunction, meaning that the total uncertainty on the ephemeris grows much more slowly with time. Our results allow transit times predicted to within 30 minutes with 3-sigma confidence through at least 2030, except for K2-167 b which has a transit uncertainty of 1.1 hours in 2030 (as compared to the 40.6 hour uncertainty the discovery ephemeris would provide, see Table \ref{tab:comparison} and Figure \ref{fig:JWST}). Therefore, K2-167 b would likely require additional transit follow up over the next few years to have an ephemeris precise enough to enable {\it JWST} observations near the expected end of the mission.

\section{Conclusion}
\label{sec:conclusion}
Using observations from the \ktwo\ and \TESS\ missions combined with archival spectroscopy, we reanalyzed four known \Ktwo\ planetary systems, providing updated system parameters and improved ephemerides for future follow up efforts. Additionally, we combined the known parallax for each system from the {\it Gaia} mission, allowing us to refine the stellar parameters within our global fits. We performed this case study on K2-114 b, K2-167 b, K2-237 b, and K2-261 b, and our updated ephemerides reduced the uncertainty on the orbital period by factors between 7 and 66. As a result of extending the photometric baseline for each system, we are now able to confidently predict future transit times {to within \(\sim\)1 hour through the extent of the JWST prime mission.} Additionally, \TESS\ will observe roughly half of all \Ktwo\ campaigns during the first extended mission, providing the first opportunity to perform an analysis similar to what is presented here, but on a much larger scale (hundreds of systems). This work also shows the importance of updating and maintaining accurate ephemerides, as most known exoplanets have not been reobserved, until recently by \TESS. Therefore, it is likely that many of the known planet ephemerides are, or will be, stale, limiting the ability to conduct detailed follow-up. The \tess\ discovered exoplanet ephemerides will also quickly degrade since most will only be discovered with a $\sim$27 day baseline of observations \citep{Dragomir:2020}. Continued monitoring and updating of transit ephemerides will likely be necessary to conduct future targeted follow-up observations, and this paper is part of a larger effort to reanalyze previously discovered planets using observations from new missions like \textit{Gaia} and \textit{TESS}. Future work should use the method presented here to reanalyze all known exoplanets observed by \TESS, providing the community with a larger pool of targets on which to perform detailed characterization in the era of {\it JWST}.

\software{{Lightkurve \citep{Lightkurve:2018}}, EXOFASTv2 \citep{Eastman:2013, Eastman:2017}, AstroImageJ \citep{Collins:2017}}
\facilities{\TESS, \ktwo, {Keck (\textit{HIRES}), La Silla 1.2m (\textit{CORALIE}), Nordic Optical 2.56m (\textit{FIES}), La Silla 3.6m (\textit{HARPS}), Telescopio Nazionale Galileo 3.58m (\textit{HARPSN)}}}

\acknowledgements

T.D. acknowledges support from MIT's Kavli Institute as a Kavli postdoctoral fellow. This research has made use of SAO/NASA's Astrophysics Data System Bibliographic Services. This research has made use of the SIMBAD database, operated at CDS, Strasbourg, France. This work has made use of data from the European Space Agency (ESA) mission {\it Gaia} (\url{https://www.cosmos.esa.int/gaia}), processed by the {\it Gaia} Data Processing and Analysis Consortium (DPAC, \url{https://www.cosmos.esa.int/web/gaia/dpac/consortium}). Funding for the DPAC has been provided by national institutions, in particular the institutions participating in the {\it Gaia} Multilateral Agreement. This work makes use of observations from the LCO network. This research made use of Lightkurve, a Python package for \textit{Kepler} and \TESS\ data analysis.

Funding for the \TESS\ mission is provided by NASA's Science Mission directorate. We acknowledge the use of public \TESS\ Alert data from pipelines at the \TESS\ Science Office and at the \TESS\ Science Processing Operations Center. This research has made use of the Exoplanet Follow-up Observation Program website, which is operated by the California Institute of Technology, under contract with the National Aeronautics and Space Administration under the Exoplanet Exploration Program. Resources supporting this work were provided by the NASA High-End Computing (HEC) Program through the NASA Advanced Supercomputing (NAS) Division at Ames Research Center for the production of the SPOC data products. This paper includes data collected by the \TESS\ mission, which are publicly available from the Mikulski Archive for Space Telescopes (MAST).

\clearpage
\bibliographystyle{apj}
\bibliography{refs}

\end{document}